\documentclass[journal]{IEEEtran}
\IEEEoverridecommandlockouts
\usepackage{booktabs}
\usepackage{siunitx}
\usepackage{cite}
\usepackage{amsmath,amssymb,amsfonts}
\usepackage{graphics, framed}
\usepackage{graphicx}
\usepackage{epsfig}
\usepackage{epstopdf}
\usepackage{url}
\usepackage{multimedia}
\usepackage{paralist}
\usepackage[printonlyused]{acronym}
\usepackage{units}
\usepackage{algorithmic}
\usepackage{textcomp}
\usepackage{float}
\usepackage{tabularx}
\usepackage{balance}
\usepackage{subfigure}
\usepackage{xcolor}
\def\BibTeX{{\rm B\kern-.05em{\sc i\kern-.025em b}\kern-.08em
		T\kern-.1667em\lower.7ex\hbox{E}\kern-.125emX}}
\usepackage{mathtools}
\usepackage{cuted} 
\usepackage[export]{adjustbox} 

\usepackage{dblfloatfix}
\usepackage{lipsum}
\usepackage{color,soul}
\usepackage{multirow}
\usepackage[overload]{textcase}

\usepackage{color, soul}
\sethlcolor{white}
\makeatletter
\def\SOUL@hlpreamble{%
	\setul{\dp\strutbox}{\dimexpr\ht\strutbox+\dp\strutbox\relax}%
	\let\SOUL@stcolor\SOUL@hlcolor
	\SOUL@stpreamble
}

\makeatother
\newcommand{\FGR}[1]{Fig.~\ref{#1}}

\newcommand{\SEC}[1]{Section~\ref{#1}}
\newcommand{\TAB}[1]{Table~\ref{#1}}
\def\rec{r(t)}
\def\trx{x(t)}
\def\awgn{\omega(t)}
\def\ch{\rho(t)}

\acrodef{CFAR}[CFAR]{constant false alarm rate}
\acrodef{FAM}[FAM]{\ac{FFT} accumulation method}
\acrodef{CAF}[CAF]{cyclic autocorrelation function}
\acrodef{SCD}[SCD]{spectral correlation density}
\acrodef{SCF}[SCF]{spectral correlation function}
\acrodef{SVM}[SVM]{support vector machine}
\acrodef{PSD}[PSD]{power spectral density}
\acrodef{AWGN}[AWGN]{additive white Gaussian noise}
\acrodef{iff}[iff]{if and only if}
\acrodef{CR}[CR]{cognitive radio}
\acrodef{SNR}[SNR]{signal-to-noise ratio}
\acrodef{NLOS}[NLOS]{non line–-of–-sight}
\acrodef{LOS}[LOS]{line–-of–-sight}
\acrodef{ADAM}[ADAM]{adaptive moment estimation}
\acrodef{GSM}[GSM]{Global System for Mobile communications}
\acrodef{WCDMA}[WCDMA]{wideband code division multiple access}
\acrodef{LTE}[LTE]{Long-Term Evolution}
\acrodef{CPU}[CPU]{central processing unit}
\acrodef{GPU}[GPU]{graphics processing unit}
\acrodef{AP}[AP]{amplitude-phase representation}
\acrodef{ISM}[ISM]{industrial scientific medical}

\acrodef{5G}[5G]{5\textsuperscript{th}-Generation}
\acrodef{BW}[BW]{bandwidth}
\acrodef{CW}[CW]{continuous wave}
\acrodef{D2D}[D2D]{device-to-device}
\acrodef{dB}[dB]{decibel}
\acrodef{dBi}[dBi]{decibel isotropic}
\acrodef{dBm}[dBm]{decibel over a milliwatt}
\acrodef{Gbps}[Gbps]{gigabit per second}
\acrodef{GHz}[GHz]{gigahertz}
\acrodef{Hz}[Hz]{hertz}
\acrodef{IF}[IF]{intermediate frequency}
\acrodef{IFFT}[IFFT]{inverse fast Fourier Transform}
\acrodef{FFT}[FFT]{fast Fourier transform}
\acrodef{KHz}[KHz]{kilohertz}
\acrodef{LO}[LO]{local oscillator}
\acrodef{LOS}[LOS]{line-of-sight}
\acrodef{MHz}[MHz]{megahertz}
\acrodef{MILTAL}[M\.{I}LTAL]{millimeter wave and terahertz technologies research laboratories}
\acrodef{MIMO}[MIMO]{multiple-input multiple-output}
\acrodef{mmWave}[mmWave]{millimeter wave}
\acrodef{NGWN}[NGWN]{next generation wireless network}
\acrodef{NLOS}[NLOS]{non line-of-sight}
\acrodef{OML}[OML]{Oleson Microwave Labs}
\acrodef{OLOS}[OLOS]{optical line-of-sight}
\acrodef{PNA}[PNA]{performance network analyzer}
\acrodef{QoS}[QoS]{quality of service}
\acrodef{RF}[RF]{radio frequency}
\acrodef{spar}[s-parameter]{scattering parameters}
\acrodef{subThz}[sub-\unit{}{THz}]{sub-terahertz}
\acrodef{TUBITAK}[T\"{U}B\.{I}TAK]{Scientific and Technological Research Council of Turkey}
\acrodef{USB}[USB]{universal serial bus}
\acrodef{VNA}[VNA]{vector network analyzer}
\acrodef{AoA}[AoA]{angle of arrival}
\acrodef{MLE}[MLE]{maximum likelihood estimation}
\acrodef{RAT}[RAT]{radio access technology}
\acrodef{SCA}[SCA]{successive convex approximation}
\acrodef{MAC}[MAC]{media access control}
\acrodef{MBS}[MBS]{macro base station}
\acrodef{MiBS}[MiBS]{micro base station}
\acrodef{FBS}[FBS]{femto base station}
\acrodef{OSI}[OSI]{open systems interconnection}
\acrodef{MINLP}[MINLP]{mixed integer non-linear program}
\acrodef{MT}[MT]{mobile terminal}
\acrodef{WLAN}[WLAN]{wireless local area network}
\acrodef{WiFi}[WiFi]{wireless-fidelity}

\acrodef{CDMA}[CDMA]{code division multiple access}
\acrodef{UMTS}[UMTS]{universal mobile telecommunications system}
\acrodef{OFDM}[OFDM]{orthogonal frequency division multiplexing}
\acrodef{OFDMA}[OFDMA]{orthogonal frequency division multiple access}
\acrodef{DCF}[DCF]{distributed coordination function}
\acrodef{BS}[BS]{base station}
\acrodef{DL}[DL]{deep learning}
\acrodef{NGWN}[NGWN]{next generation wireless network}
\acrodef{HetNet}[HetNet]{heterogeneous wireless network}
\acrodef{QoS}[QoS]{quality of service}
\acrodef{BARON}[BARON]{branch-and-reduce optimization navigator}
\acrodef{GAMS}[GAMS]{general algebraic modeling system}
\acrodef{AlphaBB}[AlphaBB]{$\alpha$-Branch and Bound}
\acrodef{MIDACO}[MIDACO]{Mixed Integer Distributed Ant Colony Optimization}
\acrodef{COUENNE}[COUNNE]{Convex Over and Under ENvelopes for Nonlinear Estimation}
\acrodef{LINDOGLOBAL}[LINDOGLOBAL]{LINDO global optimization procedure}
\acrodef{SMS}[SMS]{short message services}
\acrodef{MCTCP}[MCTCP]{multiple-connection transmission control protocol}
\acrodef{iOS}[iOS]{iPhone operating system}
\acrodef{KKT}[KKT]{Karush-Kuhn-Tucker}
\acrodef{NP}[NP]{Non-deterministic Polynomial-time}
\acrodef{D2D}[D2D]{device-to-device}
\acrodef{UMTS}[UMTS]{universal mobile telecommunications system}
\acrodef{3GPP}[3GPP]{3rd generation partnership project}
\acrodef{VBR}[VBR]{variable bit rate}
\acrodef{CBR}[CBR]{constant bit rate}
\acrodef{RHS}[RHS]{right-hand side}
\acrodef{M2M}[M2M]{machine-to-machine}
\acrodef{MLP}[MLP]{multilayer perceptron}
\acrodef{CNN}[CNN]{convolutional neural network}
\acrodef{CLDNN}[CLDNN]{convolutional long short term memory fully connected deep neural network}
\acrodef{RF}[RF]{radio frequency}
\acrodef{LSTM}[LSTM]{long short term memory}
\acrodef{WFM}[WFM]{wide-band FM}
\acrodef{DVB}[DVB]{digital video broadcast}
\acrodef{I/Q}[I/Q]{in-phase/quadrature}
\acrodef{ReLU}[ReLU]{rectified linear unit}
\acrodef{t-SNE}[t-SNE]{t-distributed stochastic neighbor embedding}
\acrodef{CFD}[CFD]{cyclostationary features detection}
\acrodef{IoT}[IoT]{Internet of things}
\acrodef{ML}[ML]{machine learning}
\ifCLASSINFOpdf
\else
\fi

\begin{document}
	\title{Spectrum Sensing and Signal Identification with Deep Learning based on Spectral Correlation Function}

	\author{K{\"{u}}r{\c{s}}at Tekb{\i}y{\i}k,~\IEEEmembership{Graduate Student Member,~IEEE,} \"{O}zkan Akbunar,~\IEEEmembership{Student Member,~IEEE,}\\Ali R{\i}za Ekti,~\IEEEmembership{Senior Member,~IEEE,} Ali G\"{o}r\c{c}in,~\IEEEmembership{Senior Member,~IEEE,}\\G{\"{u}}ne{\c{s}} Karabulut Kurt,~\IEEEmembership{Senior Member,~IEEE,} Khalid A. Qaraqe,~\IEEEmembership{Senior Member,~IEEE}
		
		\thanks{K. Tekb{\i}y{\i}k, \"{O}. Akbunar, A.R. Ekti and A. G\"{o}r\c{c}in are with the Informatics and Information Security Research Center (B{\.{I}}LGEM), T{\"{U}}B{\.{I}}TAK, Kocaeli, Turkey.~~ e-mail: \{kursat.tekbiyik, ozkan.akbunar, aliriza.ekti, ali.gorcin\}@tubitak.gov.tr}
		\thanks{A.R. Ekti is with the Department of Electrical-Electronics Engineering, Bal{\i}kesir University, Bal{\i}kesir, Turkey.~~ e-mail: arekti@balikesir.edu.tr}
		\thanks{A. G\"{o}r\c{c}in is with the Department of Electronics and Communications Engineering, Y{{\i}}ld{{\i}}z Technical University, {\.{I}}stanbul, Turkey.~~ e-mail: agorcin@yildiz.edu.tr}
		\thanks{K. Tekb{\i}y{\i}k and G.K. Kurt are with the Department of Electronics and Communications Engineering, {\.{I}}stanbul Technical University, {\.{I}}stanbul, Turkey.~~ e-mail: \{tekbiyik, gkurt\}@itu.edu.tr}
		\thanks{K. A. Qaraqe is with the Department of Electrical and Computer Engineering, Texas A\&M University at Qatar, Doha, Qatar.~~ e-mail: khalid.qaraqe@qatar.tamu.edu}
		
		\thanks{This work has received funding from the European Commission Horizon 2020 research
			and innovation programme BEYOND5 under grant agreement No. 876124.\protect}
		\thanks{This study has been supported by NPRP12S-0225-190152 from the Qatar National Research Fund (a member of The Qatar Foundation).\protect}
	}
	
	\IEEEoverridecommandlockouts

	\maketitle

	\begin{abstract}
		
		Spectrum sensing is one of the means of utilizing the scarce source of wireless spectrum efficiently. In this paper, a convolutional neural network (CNN) model employing spectral correlation function (SCF) which is an effective characterization of cyclostationarity property, is proposed for wireless spectrum sensing and signal identification. The proposed method classifies wireless signals without a priori information and it is implemented in two different settings entitled \texttt{CASE1} and \texttt{CASE2}. In \texttt{CASE1}, signals are jointly sensed and classified. In \texttt{CASE2}, sensing and classification are conducted in a sequential manner. In contrary to the classical spectrum sensing techniques, the proposed CNN method does not require a statistical decision process and does not need to know the distinct features of signals beforehand. Implementation of the method on the measured over-the-air real-world signals in cellular bands indicates important performance gains when compared to the signal classifying deep learning networks available in the literature and against classical sensing methods. Even though the implementation herein is over cellular signals, the proposed approach can be extended to the detection and classification of any signal that exhibits cyclostationary features. Finally, the measurement-based dataset which is utilized to validate the method is shared for the purposes of reproduction of the results and further research and development.
	\end{abstract}
	
	\IEEEpeerreviewmaketitle
	\acresetall

	\section{Introduction}\label{sec:intro}
	
	Today's wireless communication systems have to bear an unprecedented increase in data transmission volume.  It is essential for wireless communication networks to utilize the limited source of spectrum as efficiently and effectively as possible to meet the demand~\mbox{\cite{haykin2015cognitive}}. Furthermore, the efforts including the deployment of small cells, utilizing mmWave bands, effective spectrum usage algorithms, massive \ac{MIMO} systems\cite{andrews2014will}, and cognitive radio networks target the same goal. Cognitive radios aim to attend this purpose by sharing the spectrum dynamically among users; thus, spectrum sensing and signal identification became major techniques for cognitive radio networks.Considering joint communications, sensing, and localization demanded by 6G and beyond, efficient spectrum allocation will be more crucial for heterogeneous networks. For instance, 5G NR Release 16 introduces dynamic spectrum sharing which is novel method enabling parallel operation of 5G and \ac{LTE} in the same band~\cite{de2021convergent}. Furthermore, it is envisioned that radar and communications system will share the same frequency band~\cite{wild2021joint}.

	When spectrum sensing and signal identification techniques are considered, it is seen that sensing techniques of energy detection and matched filtering require a priori information such as number of second order noise statistics, cyclic frequencies and particular pulse shaping filter characteristics to operate. Moreover, after processing of the received signals, a statistical decision mechanism should be implemented to complete the sensing process~\cite{gorcin2014signal}. Such cumbersome process can hamper the agile decision making requirements of 5G and beyond networks, thus, classical sensing paradigm can not satisfy the requirements of fast changing operation environment of contemporary and future wireless communications networks. In this context, \ac{DL} has been proposed as a solution to the parameter adaptation issues of classical techniques. This stems from the known ability of \ac{DL} techniques in extracting the intrinsic features of given inputs through a convolutional process. The use of \ac{DL} based approaches also eliminates the need for a statistical decision mechanism at the end of the identification process. Along this line, the recent study shows that \ac{DL} methods outperform classical approaches in signal detection in the spectrum \cite{lees2019deep}. To achieve the requirements for 5G and beyond wireless networks, an intelligent radio design for spectrum sensing and signal identification is required and such solution can be realized with the help of \ac{ML} algorithms~\cite{Qin202020years} utilizing features such cyclostationarity of signals~\cite{gardner1991}.
	
	\subsection{Related Work}
	When the literature on the implementation of artificial intelligence techniques for spectrum sensing and signal identification purposes are considered, it is initially seen that \acp{CNN} are trained with high-order statistics of single carrier signals for modulation classification~\cite{o2018over}. A \ac{CNN} classifier is used for modulation and interference identification for \ac{ISM} bands by utilizing \ac{FFT}, \ac{AP} and \ac{I/Q} features for training \cite{kulin2018end}. Another study \cite{kokalj2019autoencoders} focused on the protocol classification in \ac{ISM} band by utilizing fully connected neural networks. As another example of the application of \ac{DL} to signal classification, \ac{LSTM} is deployed for modulation classification and identification of \ac{DVB}, Tetra, \ac{LTE}, \ac{GSM}, \ac{WFM} signals by using \ac{AP} and \ac{FFT} magnitude for training \cite{rajendran2018deep}. The performance of the proposed model is high, however, it employs synthetic data generated from MATLAB. In the real channels, there are numerous phenomenons, which further complicate the signal characteristics.

	On the other hand, cyclostationarity signal analysis has been explored for modulation classification, parameter estimation and spectrum sensing for more than 20 years. In addition to being an established method for spectrum sensing in cognitive radio domain, \ac{CFD} is also utilized to distinguish generic modulations such as M-PSK, M-FSK, and M-QAM \cite{gardner1991, han2006spectral}. When the \ac{RAT} identification \cite{menguconer_2007} is considered, second order cyclostationarity is employed for classification of \ac{LTE} and \ac{GSM} signals \cite{karami2015identification}. Later, a tree-based classification approach is proposed to identify \ac{GSM}, cdma2000, \ac{UMTS} and \ac{LTE} signals \cite{eldemerdash2017identification}. 
	
	\subsection{The Contributions}
	
	\subsubsection{Methodological novelty}
	\ac{CFD} depends on extracting the underlying features using likelihood-based techniques utilizing statistical decision mechanisms and for \ac{CFD} to operate under the dynamically changing communication medium, an additional mechanism to adaptively adjust decision parameters such as thresholds and the number samples is required \cite{hazza2013overview}. On the other hand, even though employment of \ac{DL} techniques for the purposes of spectrum sensing and signal identification implies considerable advantages in terms of performance and complexity, utilization of \ac{FFT}, \ac{AP} and \ac{I/Q} as input features to the intelligent networks do not lead stable and dependable results due to the rapidly and significantly changing wireless communications medium between the nodes. Therefore, \textit{this study proposes application of SCF as input feature to CNNs for blind wireless signal identification}. The problems of spectrum sensing and signal identification are framed into two particular contexts which utilize a novel \ac{CNN} model designed and trained with \ac{SCF} of wireless signals without bi-frequency mapping. Therefore\textcolor{black}{,} the proposed method can be employed either to decide whether the signal is present or not in the spectrum or to distinguish signals from each other. Sensing and identification performance of the method is tested and validated utilizing real-life over-the-air signal measurements of \mbox{\ac{GSM}}, \mbox{\ac{UMTS}}, and \mbox{\ac{LTE}} signals.

	\hl{The proposed method approaches to the problems of sensing and identification from the aspects of two cases; in \texttt{CASE1}, the designed \mbox{\ac{CNN}} model is fed directly with the \mbox{\ac{SCF}}s of measurements of \mbox{\ac{GSM}}, \mbox{\ac{UMTS}}, \mbox{\ac{LTE}} along with SCF of spectrum which is only comprised of noise. Sensing and classification are executed jointly for \texttt{CASE1}. On the other hand in \texttt{CASE2}, a two-step approach is adopted; first, as a spectrum sensing method to measure the spectrum occupancy is conducted and this stage is followed by a signal classification procedure.}
	
	\subsubsection{Novelty in terms of numerical studies}
	In terms of performance analysis, first, a comparative analysis is conducted and superiority of \mbox{\ac{SCF}} over the features of \ac{I/Q}, \mbox{\ac{AP}} and \mbox{\ac{FFT}} is shown for the purpose of training of \ac{DL} networks. Second, comparison with the existing \mbox{\ac{DL}} methods such as \ac{CLDNN}~\cite{ramjee2019fast}, \ac{LSTM}~\cite{rajendran2018deep}, DenseNet~\cite{huang2017densely}, ResNet~\cite{o2018over} are given in terms of accuracy, memory consumption and computational complexity. Third, it is shown that the proposed  method outperforms \acp{SVM} trained with \ac{SCF}, which is our previous study. Fourth, the performance of the proposed method is compared with the classical spectrum sensing technique of \ac{CFD}, which requires the cyclic frequencies as a priori information. The identification results indicate important performance improvements over the aforementioned techniques. 
	
	\subsubsection{Novelty in terms of experimental activities} Focusing on the valuable information in the dataset is an important metric for the proposed method; thus, it is denoted that utilizing only the meaningful part of the input matrices improves the classification performance along with alleviation in training time and complexity. On the other hand, the general dataset, which has been developed from measurements taken through a comprehensive measurement campaign conducted in different locations and frequency bands, is shared publicly in~\cite{safrgh5920}. Therefore\textcolor{black}{,} the measurement-based dataset is open to researchers as a comprehensive resource in the development and validation of their work.
	
	\subsubsection{Applicability for future research problems}
	Even though in this work the scope of implementation is focused on cellular signals, the introduced identification system can be directly used for detection and classification of any signal that exhibit cyclostationary features. All the analyses are based on the real-world measurements taken during an extensive measurement campaign conducted at different locations with varying environmental conditions such as channel fading statistics and \mbox{\ac{SNR}} levels. Finally, the measurement data that this work is experimented on is also shared for reproducibility of this work and to support future research and development activities in this domain.
	
	\subsection{Organization of the Paper}
	The rest of the paper is structured as follows. Background information on the system model, cyclostationary analysis and \acp{CNN} is presented in \SEC{sec:background}.  The problem statement is given in \SEC{sec:problem_statement}. The proposed \ac{CNN} model is described in \SEC{sec:proposed_cnn}. The details of the measurement setup and dataset utilized in this study are given in \SEC{sec:dataset_generation}. \SEC{sec:performance} presents measurement results and details the classification performance of the proposed method. The concluding remarks are provided in \SEC{sec:conclusion}.
	
	\section{Background}\label{sec:background}
	Assuming that received signal is down converted to baseband before further processing, first the complex baseband equivalent of the received signal, $\rec$ should be defined. When the presence of fading environment with thermal noise, received signal can be given as
	\begin{align}
	\rec = \rho(t)*x(t) + \awgn,
	\end{align}
	\noindent where $\awgn$ denotes the complex \ac{AWGN} with $\mathcal{C}\mathcal{N}(0,\sigma^2_N)$ in the form of $\awgn=\omega_I(t)+j\omega_Q(t)$ as both $\omega_I(t)$ and $\omega_Q(t)$ being $\mathcal{N}(0,{\sigma^2_N}/2)$ and $j=\sqrt{-1}$; the complex baseband equivalent of the transmitted signals is denoted as $\trx$; and $\ch$ stands for the impulse response for the time-invariant wireless channel because of extremely short observation time for a signal.
	
	Depending on the idle or busy state of the mobile propagation channel in the \ac{RF} spectrum, the signal detection process of deep learning methods can be modelled as a binary hypothesis test
	
	\begin{equation}
	\label{eq_signal_model_sensing}
	\rec =  \left\{\begin{matrix} \rho(t)x(t) + \awgn,~~\hfill H_1\\ \awgn,~~\hfill H_0. \end{matrix}\right.
	\end{equation} 
	\noindent $H_0$ and $H_1$ hypotheses stand for the presence of noise only and the unknown signal, respectively. Therefore, the problem statement can be stated as identification of the presence of the unknown signal, $x(t)$, and classification of the $x(t)$.
	
	\subsection{Cyclostationarity}\label{sec:cyclostationarity}
	
	Cyclostationary signal processing leads to extracting hidden periodicities in a received signal, $\rec$. Since these periodicities (\textit{e.g.}, symbol periods, spreading codes, and guard intervals) exhibit unique characteristics for different signals, they provide the necessary information for identification. Thus, the unknown signals $\trx$ can be identified by using cyclostationary features to obtain the statistical characteristics of $\rec$ in the presence of $\awgn$ and multipath fading without a priori information. A nonlinear transformation, second-order cyclostationarity of a signal can be expressed as
	\begin{align}
	s_{\tau}(t)=\mathbb{E}\left\{r(t+\tau/2)r^{*}(t-\tau/2)\right\},
	\label{1}
	\end{align}
	\noindent where $\mathrm{s}_{\tau}(t)$ is the autocorrelation of $\rec$. Assuming that the autocorrelation function is periodic with $T_{0}$ for second-order cyclostationary signals, a Fourier series expansion of $s_{\tau}(t)$ is given as
	
	\begin{align}
	\mathbb{R}_{r}^{\alpha}\left(\tau\right) = \frac{1}{T_0}\int_{-T_0/2}^{T_0/2} s_{\tau}(t)e^{-j2\pi\alpha t}dt,
	\end{align} 
	\noindent where $\mathbb{R}_{r}^{\alpha}\left(\tau\right)$ is the \ac{CAF} and $\alpha$ values denote the cyclic frequencies. 
	
	The Fourier transform of the \ac{CAF} for a fixed $\alpha$ is given with the cyclic Wiener relation \cite{gardner1991}
	\begin{align}
	\mathbb{S}_r(f) = \int_{-T/2}^{T/2} \mathbb{R}_{r}^{\alpha}\left(\tau\right) e^{-j2\pi f \tau}d\tau,
	\end{align}
	\noindent where $\mathbb{S}_r(f)$ is called as \ac{SCF} which is equal to the \ac{PSD} when $\alpha$ is zero. 
	
	The computational complexity of calculating \ac{SCF} is relatively high. However, this complexity can be decreased by using the \ac{FAM} based on time smoothing via \ac{FFT} \cite{roberts1991}. \ac{FAM} estimates the \ac{SCF} as
	
	\begin{align}
	\label{eq:SCF_FAM}
	\mathbb{S}_{r_{T}} &= \sum_{k} R_T(kL, f)R_T^*(kL, f)g_{c}(n-k)e^{-i2\pi kq/P},
	\end{align}
	\noindent where $R_{T}(n, f)$ denotes the complex demodulates which is the $N'$-point \ac{FFT} of $r(n)$ passed through a Hamming window and can be computed by
	\begin{align}
	R_{T}(n, f) &= \sum_{k=-N'/2}^{N'/2}a(k)r(n-k)e^{-i2\pi f(n-k)T_{s}},
	\end{align}
	%%%%%%%%%%%%%%%%%%%%%%%%%%%%%%%%%%%%%%%%%%%%%%%%%%%%%NOTE END
	
	\noindent where $a(n)$ and $g_c(n)$ are both data tapering windows. The symbols $N'$, $T_s$, and $L$ denote the channelization length, sampling period, and sample size of hopping blocks, respectively. The ratio between the number of total samples and $L$ is employed as the length of second \ac{FFT}, whose length is denoted as $P$. The \ac{FAM} has six implementation steps. These steps are respectively channelization, windowing, $N'$-point \ac{FFT}, complex multiplication, $P$-point \ac{FFT} and bi-frequency mapping. In the study, the unit rectangle and Hamming windows are employed as $g_c(n)$ and $a(n)$, respectively. \FGR{fig:SCF3} illustrates \acp{SCF} results in bi-frequency plane, which are estimated by \ac{FAM} algorithm for \ac{GSM}, \ac{UMTS}, and \ac{LTE} along with the noise. Consequently, the input matrix, $\bold{X}_k^{SCF}$, to be fed into classifier model is given as
	\begin{align}
	\bold{X}_k^{SCF} = |\mathbb{S}_{r_{T}}^{}(nL, f)|,
	\label{eq7}
	\end{align}
	
	\begin{figure*}[!t]
		\centering
		\subfigure[AWGN]{%
			\label{fig:awgn}%
			\includegraphics[width=0.22\textwidth]{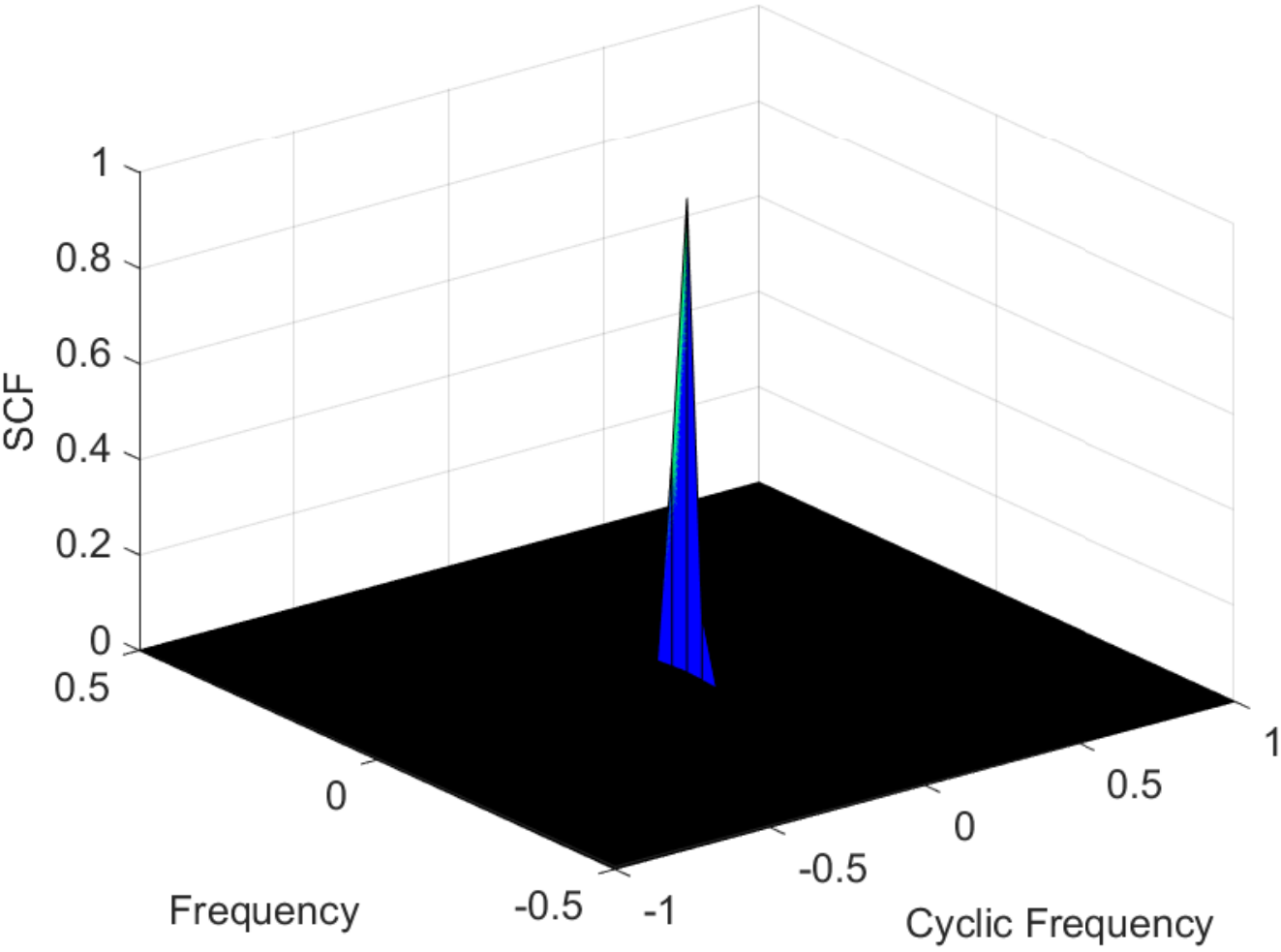}}
		\quad
		\subfigure[GSM]{%
			\label{fig:GSM}%
			\includegraphics[width=0.22\textwidth]{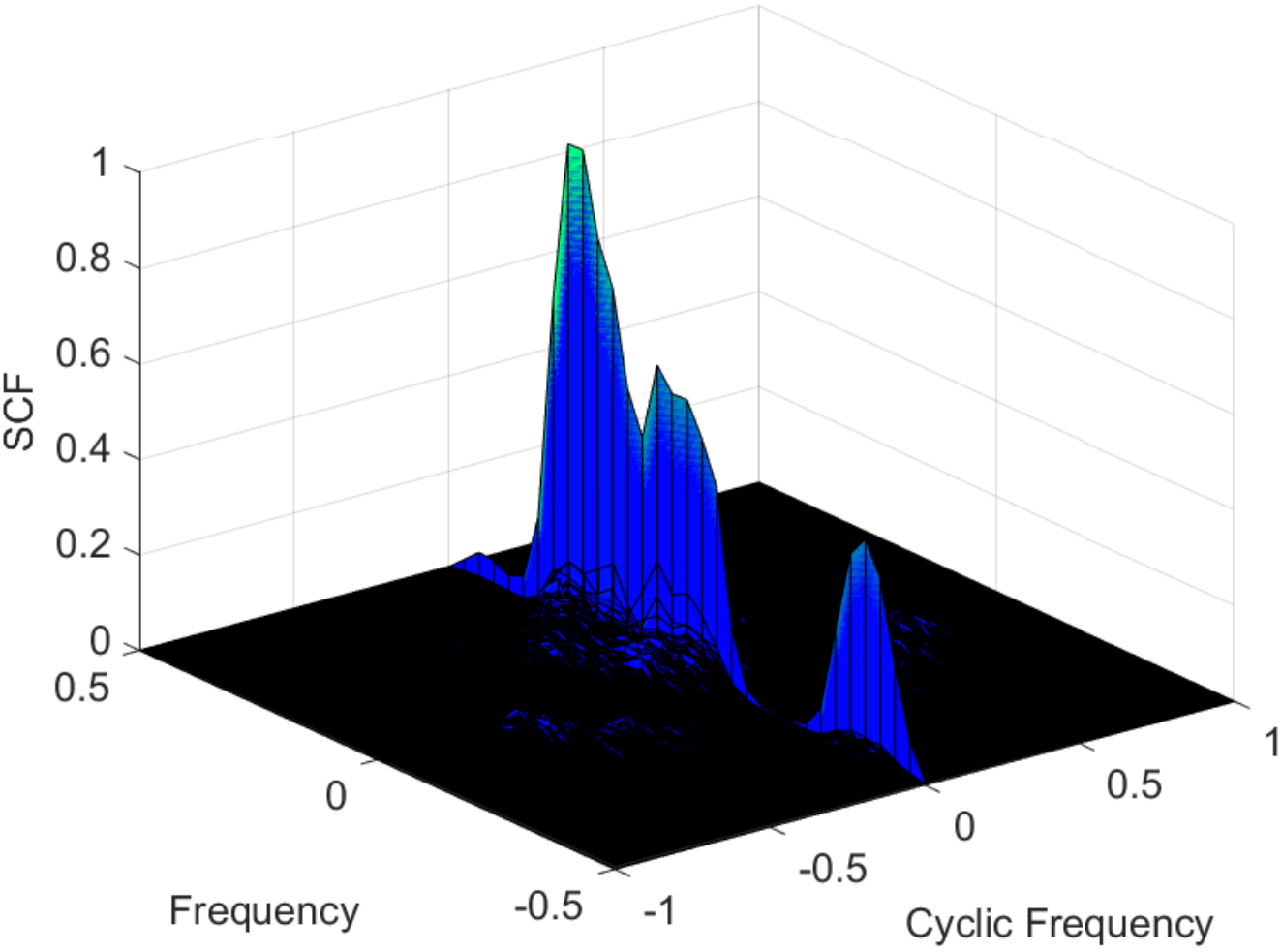}}
		\quad
		\subfigure[UMTS]{%
			\label{fig:UMTS}%
			\includegraphics[width=0.22\textwidth]{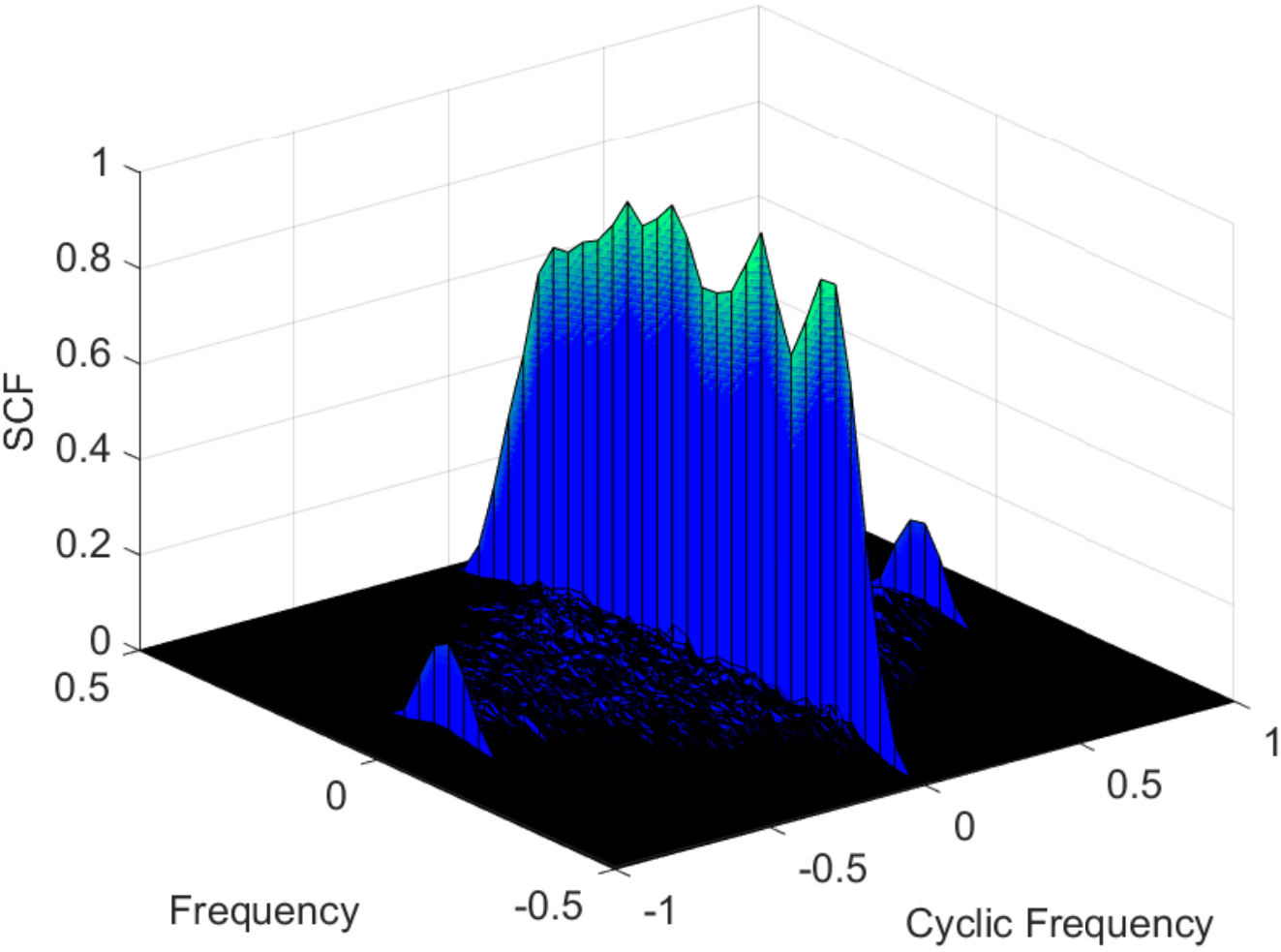}}%
		\quad
		\subfigure[LTE]{%
			\label{fig:LTE}%
			\includegraphics[width=0.22\textwidth]{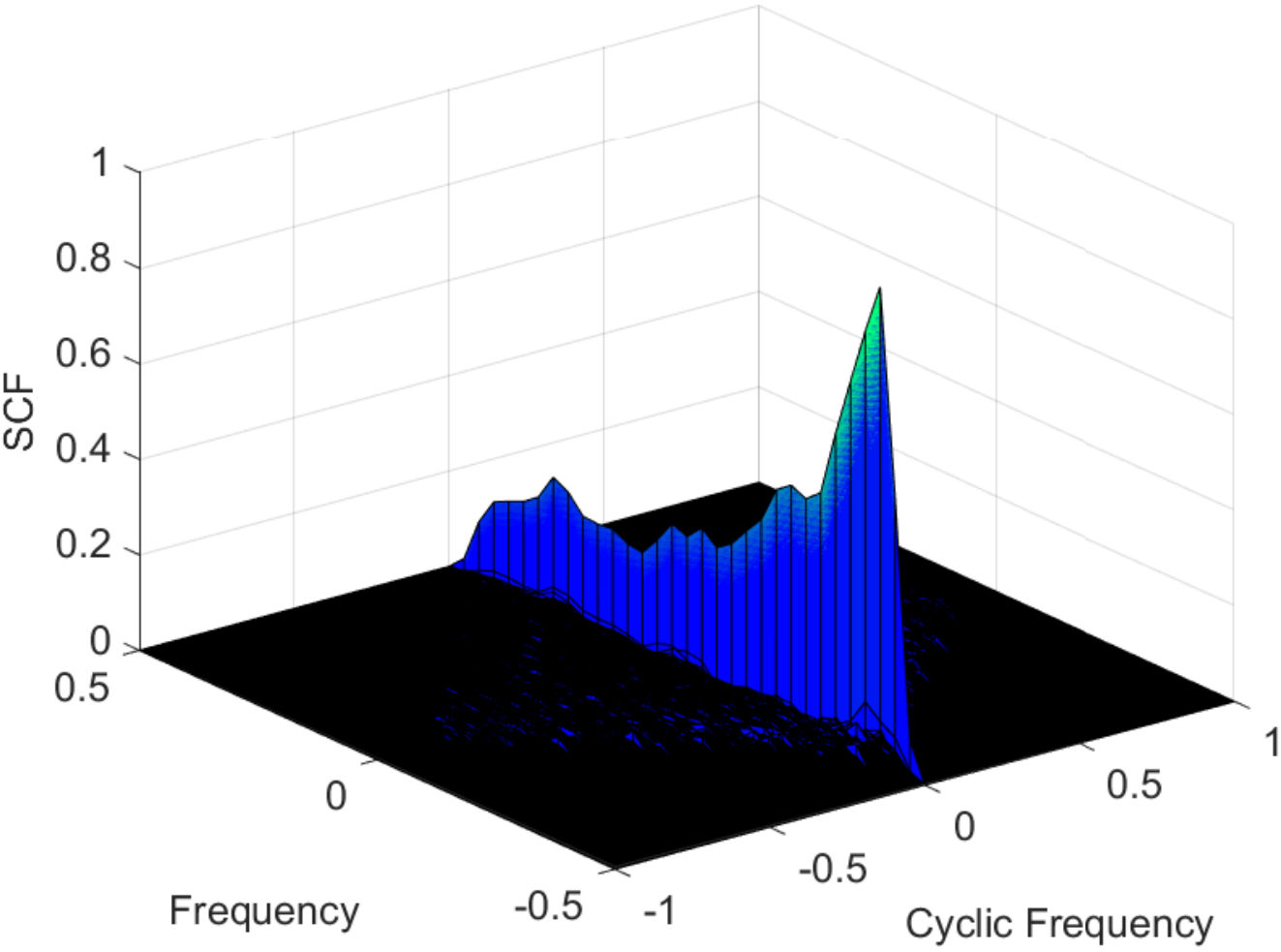}}%
		\caption{\hl{\mbox{\ac{FAM}} based SCF estimates of cellular signals in bi-frequency plane. It is easily observed that the signals show different cyclic characteristics. The noise does not show cyclic characteristics as SCF of noise gives only peak at the center of bi-frequency plane where the cyclic frequency is zero.}}
		\label{fig:SCF3}
	\end{figure*}

	\subsection{Amplitude-Phase}
	The amplitude and phase values of time-domain \ac{I/Q} data can be used to establish a real-valued classification feature matrix, $\bold{X}_k^{AP}$. This feature matrix is composed of the amplitude and phase vectors of the received signal samples. So, $\bold{X}_k^{AP}$ is defined as 
	\begin{align}
	\bold{X}_k^{AP}=\begin{bmatrix} \bold{x}_A^{T} \\
	\bold{x}_{\phi}^{T}
	\end{bmatrix},
	\end{align}
	\noindent where $\bold{x}_A=({r_q}^2+{r_i}^2)^{\frac{1}{2}}$ and $\bold{x}_{\phi}=\arctan(\frac{r_q}{r_i})$ denote the amplitude and phase vectors, respectively.
	
	\subsection{Fast Fourier Transform}
	The characteristics of signals in frequency domain can be employed as discriminating classification features. The \ac{FFT} of the received signal is used to obtain a real-valued classification feature matrix $\bold{X}_k^{FFT}$ as 
	\begin{align}
	\bold{f}&=\mathcal{F}(r),~
	\bold{X}_k^{FFT}=\begin{bmatrix} \bold{f}_{re}^{T}  \\
	\bold{f}_{im}^{T}
	\end{bmatrix}, 
	\end{align}
	\noindent where $\mathcal{F}(\cdot)$ stands for the \ac{FFT} of the received signals; $\bold{f}_{re}$ and $\bold{f}_{im}$ are real and imaginary parts of $\bold{f}$, respectively. 
	
	\subsection{The Convolutional Neural Networks}\label{sec:cnn}
	\ac{CNN} is a class of deep neural networks which is mainly employed in image classification and recognition. \textcolor{black}{\ac{CNN} process inputs like a visual system in human. In other words, it extracts features in an input rather than fitting data~\cite{zhang2017convolutional}. In this study, we utilize the input matrices which resemble image consisting of features in a specific positions as seen in~\FGR{fig:SCF3}.} Still, it has been recently extended to several application areas. \acp{CNN} have two stages: feature extraction and classification. In feature extraction, a convolutional layer is followed by a pooling layer. In the convolution layer, the feature matrix is convolved with different filters to obtain convolved feature map as follows
	\begin{align}
	\mathrm{h}[i, j]=\sum_{p=1}^{m} \sum_{l=1}^{n} w_{p, l} \bold{X_k}[{i+p-1, j+l-1}],
	\end{align}
	where $w_{p, l}$ is the element at $p$-th row and $l$-th column of the $m \times n$ filter matrix, and $\bold{X_k}\left[\cdot,\cdot\right]$ denotes the elements of feature matrix convolved by $w_{p, l}$. The convolution layer is followed by the pooling layer to reduce computational complexity and training time, and control over-fitting due to the fact that pooling layer makes the activation less sensitive to feature locations \cite{zeiler2013stochastic}. The $u \times v$ maximum pooling operation is described as
	\begin{align}
	\mathrm{g}[i, j] =&\max \left\{h[i+a-1, j+b-1]\right\},
	\end{align}
	\noindent where $1 \leq a \leq u \text { and } 1 \leq b \leq v$. The output of the pooling layer is a 3-D tensor. This output is then reshaped into a 1-D vector. This vector is fed to the dense (fully-connected) layers for the final classification decision. 
	
	\section{Problem Statement}\label{sec:problem_statement}
	
	The dynamic communications environment of next generation wireless networks require fast, robust and adaptive sensing and identification of the multi-dimensional communications medium to utilize the resources quickly and efficiently ~\cite{gorcin2014signal}. In this context, spectrum sensing and signal identification becomes important means of achieving effective resource utilization. To that end, we approach the problems of sensing and identification via DL from two aspects:
	
	\texttt{CASE1}: In this case, first a novel \ac{CNN} classifier is trained with all possible classes, in this case \ac{GSM}, \ac{UMTS}, \ac{LTE} and empty spectrum which can be referred to as \ac{AWGN} only. For each signal the cyclic spectrum is constructed based on the procedures described in \SEC{sec:cyclostationarity}. The cyclic spectrum is then fed to the \ac{CNN} classifier, which is  trained with four possible inputs beforehand. Finally, the classification is made.
	
	\texttt{CASE2}: In this case a two-stage approach is adopted; at the first stage a \ac{CNN} detector (the same \ac{CNN} model defined is employed for both detection and classification for the sake of simplicity) is utilized to decide whether a signal exists in the given band or not by training the \ac{CNN} by two classes, first comprised of \ac{GSM}, \ac{UMTS}, and \ac{LTE} signals and second part with \ac{AWGN} only. Thus, in the first stage a decision is made about whether a signal exists in the spectrum or not as in the case of classical spectrum sensing. If the decision is made that there is an information bearing signal in the given band, second stage is activated utilizing a \ac{CNN} classifier, which is trained in our case with three classes (\textit{i.e.}, \ac{GSM}, \ac{UMTS}, and \ac{LTE}) and finally a decision is made for the class of the signal occupying the spectrum.
	
	Please note that the classification refers to identification of the signals, and  at the detection part of the approach $H_1$ and $H_0$ refers to the existence and non-existence of a signal over the spectrum based on binary hypothesis testing. Both \texttt{CASE1} and \texttt{CASE2} are illustrated in \FGR{fig:case1_case2}.
	
	Firstly, we can define the accuracy for \texttt{CASE1}, $P_{\texttt{CASE1}}$ as:
	\begin{align}
	P_{\texttt{CASE1}} = P(\hat{\mathbf{\chi}}_{k} = \mathbf{\chi}_{k}), \; k = 0, 1, 2, 3,
	\end{align}
	where $\mathbf{\chi}_{k}$ denotes the label array of the transmitted signals and $k$ represents the label of the classes \ac{AWGN}, \ac{GSM}, \ac{UMTS}, and \ac{LTE}, respectively. $\hat{\mathbf{\chi}}_{k}$ is array for the predicted classes of the received signals. In a short, $P_{\texttt{CASE1}}$ stands for the accuracy of four-classes classification problem. For \texttt{CASE2}, it is required to define two independent accuracy functions: the sensing accuracy, $P_{\texttt{CASE2}}^{\texttt{S}}$ and the classification accuracy, $P_{\texttt{CASE2}}^{\texttt{C}}$, which are defined as
	\begin{align}
	P_{\texttt{CASE2}}^{\texttt{S}} &= P(\hat{\mathbf{\chi}}_{\texttt{S}} = 1 |H_{1}) + P(\hat{\mathbf{\chi}}_{\texttt{S}} = 0|H_{0}),\\
	P_{\texttt{CASE2}}^{\texttt{C}} &= P(\hat{\mathbf{\chi}}_{k} = \mathbf{\chi}_{k}|H_{1}), \; k = 1, 2, 3.
	\end{align}
	$\hat{\mathbf{\chi}}_{\texttt{S}}$ is the prediction of $\mathbf{\chi}_{\texttt{S}}$ regarding to the presence of a signal in the spectrum. $\mathbf{\chi}_{k}$ stands for the predictions for the classification part of $\texttt{CASE2}$. $\mathbf{\chi}_{\texttt{S}}$ is defined for the transmitted signal as:
	\begin{align}
	\mathbf{\chi}_{\texttt{S}} = \left\{\begin{matrix} 0,~~\hfill k = 0,\\ 1,~~\hfill k = 1,2,3. \end{matrix}\right.
	\end{align}
	The overall accuracy for \texttt{CASE2} can be introduced in terms of $P_{\texttt{CASE2}}^{\texttt{S}}$ and $P_{\texttt{CASE2}}^{\texttt{C}}$ by
	\begin{align}
	P_{\texttt{CASE2}} =  P_{\texttt{CASE2}}^{\texttt{S}}P_{\texttt{CASE2}}^{\texttt{C}}.
	\end{align}
	
	\section{The Proposed CNN Model}\label{sec:proposed_cnn}
	%%%%%%%%%%%%%%%%%%%%%%%%%%%%% case1_and_case2 %%%%%%%%%%%%%%%%%%%%%%%
	\begin{figure}[!t]
		\centering
		\includegraphics[width = \linewidth]{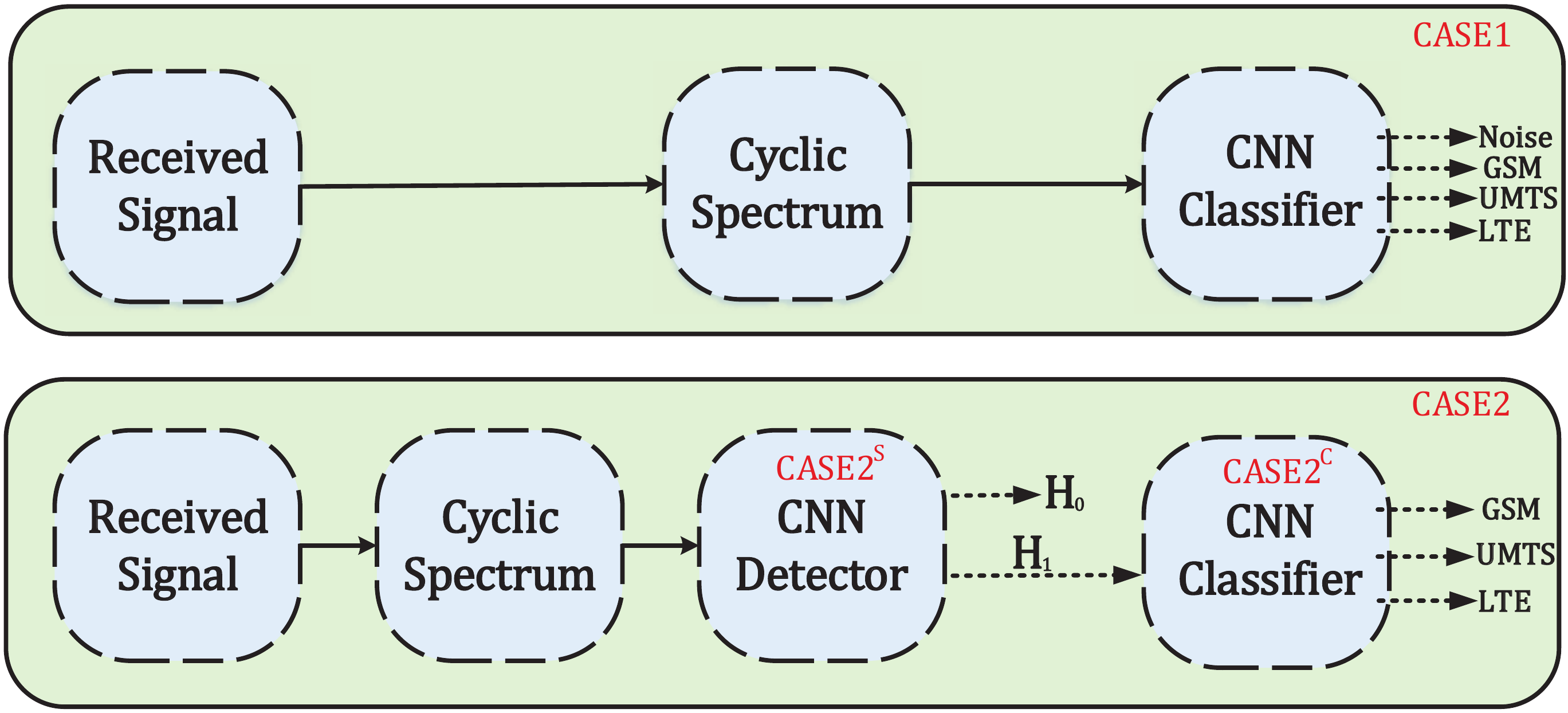}
		\caption{Two different approaches for the sensing and classification of signals. In \texttt{CASE1}, signal sensing and signal classification are jointly conducted. However, \texttt{CASE2} firstly sense signal in the spectrum, then classify.}
		\label{fig:case1_case2}
	\end{figure}
	%%%%%%%%%%%%%%%%%%%%%%%%%%%%% case1_and_case2 %%%%%%%%%%%%%%%%%%%%%%%
	
	%%%%%%%%%%%%%%%%%%%%%%%%%%%%% cnn_model_layout_table %%%%%%%%%%%%%%%%%%%%%%%
	\begin{table}[!t]
		\centering
		\caption{The proposed \ac{CNN} layout.}
		\begin{tabular}{c c}
			\toprule
			Layer & Output Dimensions \\ \midrule
			Input                  & $8193\times16$ \\         
			Conv1                  & $8193\times16\times64$ \\
			Leaky ReLU1             & $8193\times16\times64$ \\
			Max\_Pool1             & $4097\times8\times64$  \\
			Conv2                  & $4097\times8\times128$ \\
			Leaky ReLU2                  & $4097\times8\times128$ \\
			Max\_Pool2                & $2049\times4\times128$ \\
			Conv3                  & $2049\times4\times64$    \\
			Leaky ReLU3                  & $2049\times4\times64$    \\
			Max\_Pool3            & $1025\times2\times64$  \\
			Flatten                 & $131200$    \\
			Dense1                & $256$ \\
			Dense2                 & $4$   \\
			Trainable Par.         & $33,736,772$ \\  \bottomrule
		\end{tabular}
		\label{tab:cnn_layout}
	\end{table}
	%%%%%%%%%%%%%%%%%%%%%%%%%%%%% cnn_model_layout_table %%%%%%%%%%%%%%%%%%%%%%%

	As indicated in \SEC{sec:problem_statement}, the proposed method relies on a novel CNN model. Design and implementation of \ac{CNN} for classification of wireless mobile communication signals is conducted via an open source machine learning library, Keras \cite{chollet2015keras}. The proposed \ac{CNN} model consists of three convolution and three pooling layers sequentially. \hl{The convolution layers have respectively $64$, $128$, and $64$ filters. The network is terminated by two fully connected layers. First hidden layer includes 256 neurons. Second hidden layer consists of $4$ and $3$ neurons for \texttt{CASE1} and \texttt{CASE2}, respectively.} The leaky \ac{ReLU} activation function with an alpha value $0.1$ is used in each convolution layer to extract discriminating features. Leaky \ac{ReLU} is selected instead of \ac{ReLU}. Unlike \ac{ReLU}, leaky \ac{ReLU} maps larger negative values to smaller ones by a mapping line with a small slope. In each convolution layer, $3\times3$ filters are used. $2\times2$ max pooling is used to reduce the dimension and training time. A fully connected layer is formed by $256$ neurons and Leaky \ac{ReLU} activation function. Following the fully connected layers, the probabilities for each class are computed by the softmax activation function. In addition, the \ac{ADAM} optimizer is utilized when determining the model parameters. \hl{In the training phase, early stopping is employed to prevent the model from over-fitting. The patience is chosen as $10$ epochs for early stopping function and validation loss is monitored during the training. If the validation loss converges a level and remain at this level during $10$ epochs, the training is terminated and the weights at the end of training are used in the test.} The implementation layout for the proposed CNN model is given in Table~\ref{tab:cnn_layout}. The input matrices, $\bold{X}_k^{AP}$, $\bold{X}_k^{FFT}$, and $\bold{X}_k^{SCF}$ are used at the beginning of the proposed model by convolving with filters. The overall block diagram for the proposed \ac{CNN} model is depicted in \FGR{fig:cnn_model}. 
	
	When the motivation behind designing such a CNN model is considered, it should be noted firstly that the information about changes in the local regions of the mapped output is extracted by using $3\times3\times64$ filters in the first convolution layer. In this problem, because the SCF creates local differences in frequency and cyclic frequency regions, the smaller filter size is preferred to catch peaks in the feature matrices. Thus, local differences are taken into account along the layers. After the first layer determines the cyclic characteristics of all local terms as a general process, the second layer examines the properties such as location and size related to these characteristics. Here, it is aimed to deal with cyclic features in detail by increasing the number of filters to $128$. In the last layer, all properties are converted to an average of all information gathered and eventually sent to the decisive layer which is dense layer. For this reason, the number of filters in the last layer should be chosen so that sufficient information is obtained without overfitting. Therefore, the number of filters is selected as $64$ in the last layer.
	%%%%%%%%%%%%%%%%%%%%%%%%%%%%% cnn_model_layout_figure %%%%%%%%%%%%%%%%%%%%%%%
	\begin{figure}[!t]
		\centering
		\includegraphics[width = \linewidth]{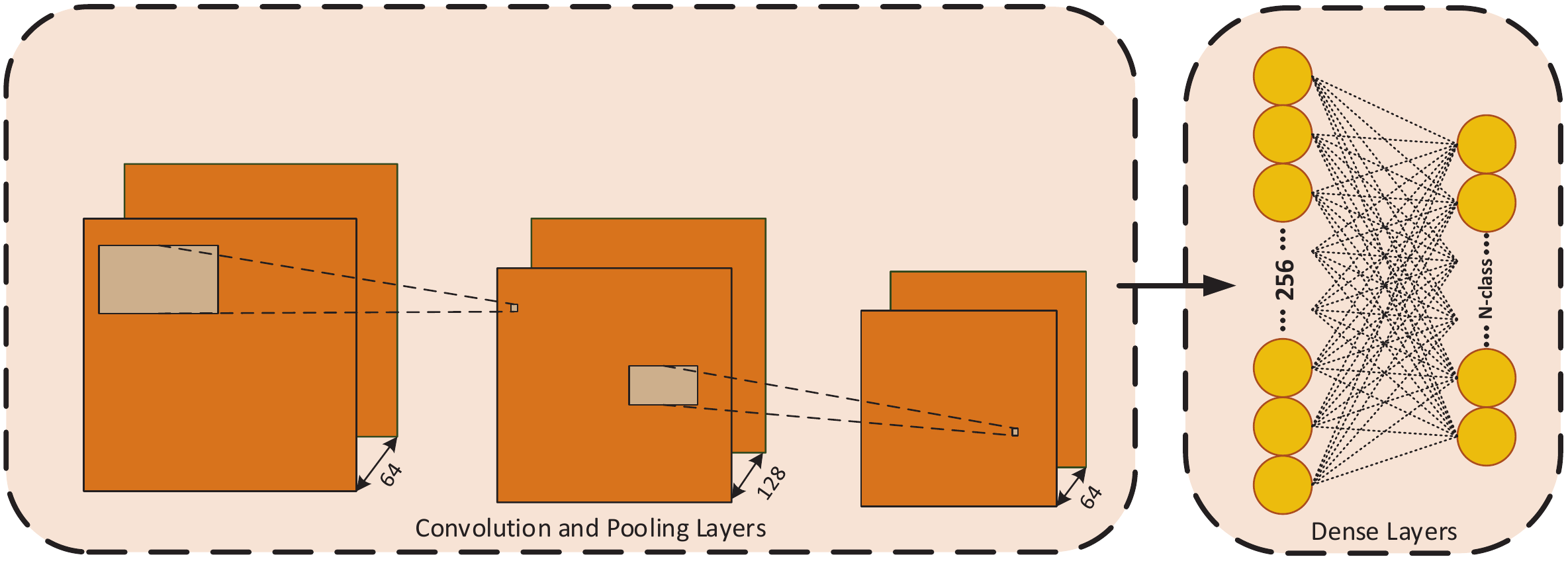}
		\caption{The proposed \ac{CNN} model consists of three convolutional layers and two dense layers with Adam optimizer with learning rate of $10^{-5}$.}
		\label{fig:cnn_model}
	\end{figure}
	%%%%%%%%%%%%%%%%%%%%%%%%%%%%% cnn_model_layout_figure %%%%%%%%%%%%%%%%%%%%%%%
	It is customary to quantify the performance of a classifier model in terms of the precision ($\Pi$), recall ($\Psi$), and $F_1$-score performance metrics. The precision metric quantifies how much positive results are actually positive, the recall provides information on how much true positives are identified correctly as positive, and $F_1$-score gives an overall measure for the accuracy of a classifier model since it is the harmonic average of precision and recall. These metrics are given as
	\begin{align}
	\Pi = \frac{\xi}{\xi+\upsilon}, \; ~~ \Psi = \frac{\xi}{\xi + \mu},\; ~~
	F_1\textrm{-score} = 2\times\frac{\Pi \times \Psi}{\Pi + \Psi},
	\label{eq:metrics}
	\end{align}
	\noindent where $\xi$, $\upsilon$, and $\mu$ denote the numbers of true positive, false positive, and false negative, respectively.

	% As aferomentioned in the previous section, the input matrices, $\bold{X}_k^{SCF}$, $\bold{X}_k^{AP}$, and $\bold{X}_k^{SCF}$, are individually fed into the proposed \ac{CNN} model.
	\section{Measurement Methodology and Dataset Generation}\label{sec:dataset_generation}
	\begin{figure*}[!t]
		\centering
		\includegraphics[width=\linewidth]{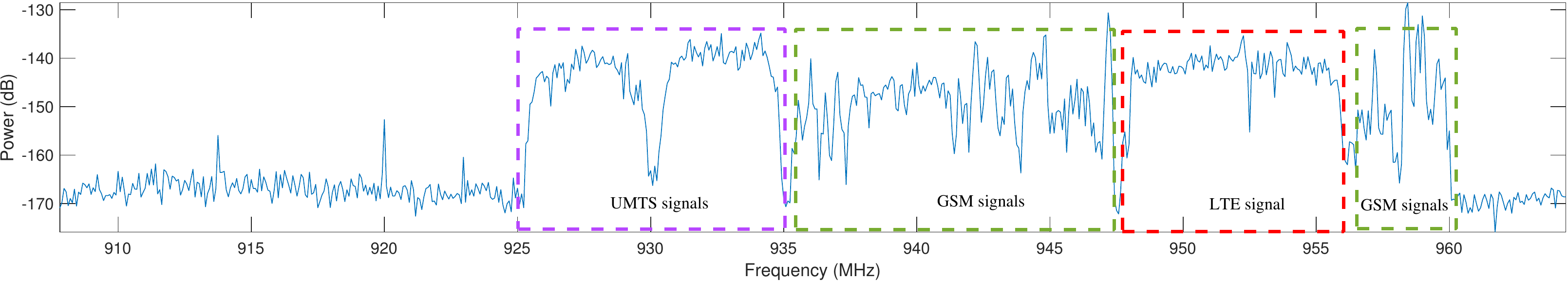}
		\caption{\hl{This snapshot of spectrum denotes a sample from the dataset comprised of cellular signals recorded during a comprehensive measurement campaign. $900$ MHz band is represented here but the measurements are not limited to that band; thus, cover all cellular bands.}}
		\label{fig:psd}
	\end{figure*}
	%%%%%%%%%%%%%%%%%%%%%%%%%%%%%%%%%%%%%%%%%%%%% Measurement Map %%%%%%%%%%%%%%%%%%%%%%%
	\begin{figure}[!t]
		\centering
		\includegraphics[width = \linewidth]{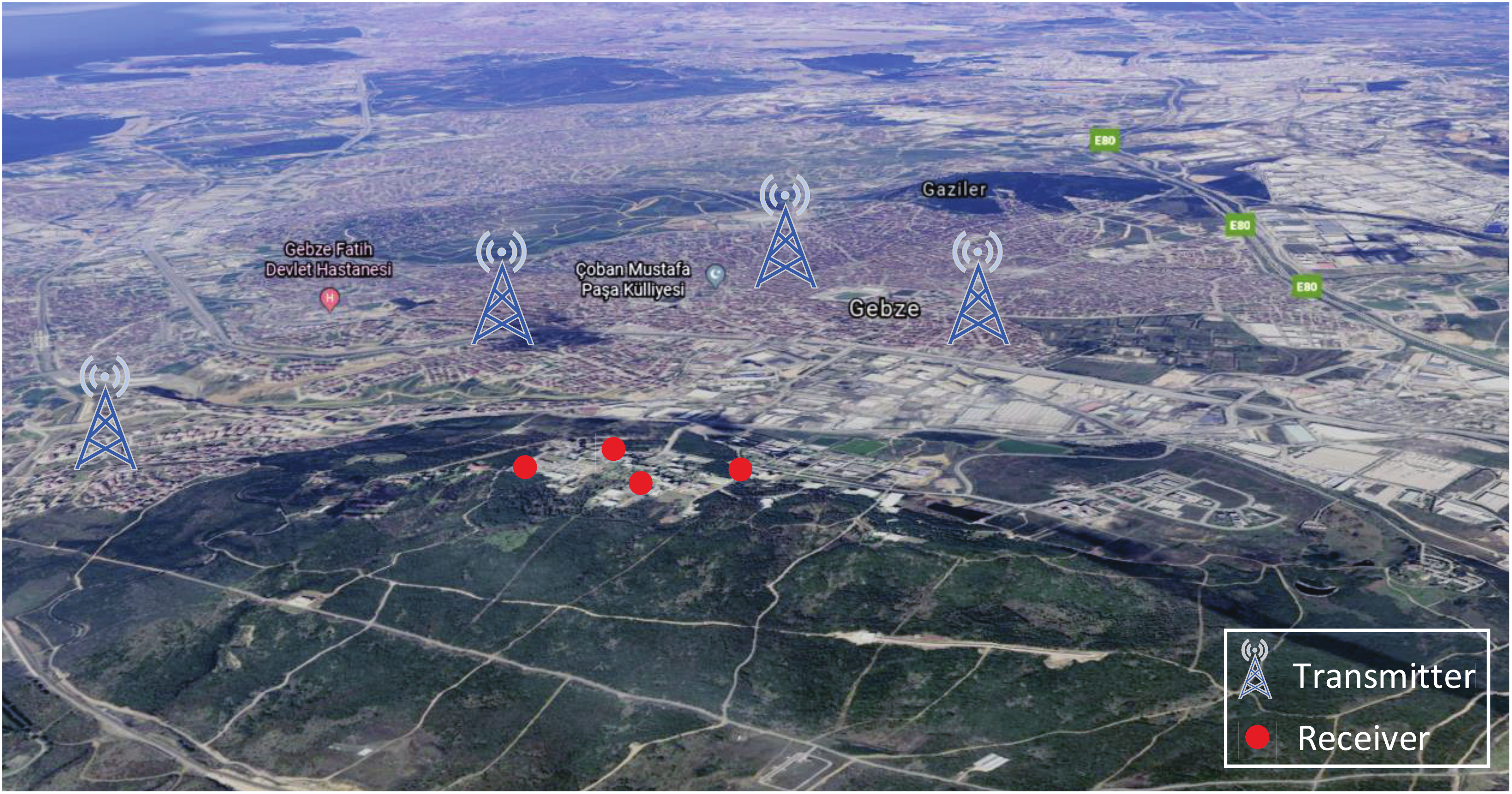}
		\caption{An overview of the measurement area. The transmitters are located in the urban area, but the receivers are in a sub-urban area.}
		\label{fig:measurement_map}
	\end{figure}
	%%%%%%%%%%%%%%%%%%%%%%%%%%%%%%%%%%%%%%%%%%%%% Measurement Map %%%%%%%%%%%%%%%%%%%%%%%

	The dataset to test and evaluate the proposed method is developed from the measurements taken through an extensive measurement campaign conducted at different locations and frequency bands. \textcolor{black}{In order to make the model robust against environmental changes, measurements have been conducted in different locations as illustrated in \FGR{fig:measurement_map}. The locations of transmitters and measurement points can be seen in \FGR{fig:measurement_map}.} It can be seen that the signals propagate through the urban area, and then reach the receivers in sub-urban area. The measurement focuses on $800$, $900$, $1800$, and $2100$ MHz frequency bands that are allocated for cellular communications. Rohde Schwarz FSW26 spectrum analyzer and a set of Yagi-Uda antennas are employed at the receiver. The measurements are unified as follows: for each signal observed in the spectrum, $16384$ \ac{I/Q} samples are taken. Measurements are conducted at $15$ different \ac{SNR} levels. Each level consists of the same number of signals which is $4000$. Therefore\textcolor{black}{,} $60000$ signals in total are recorded and included in the dataset. Sample power spectra of these signal types, obtained with the Welch's method, are shown in \FGR{fig:psd}. When the proposed method is considered, the dataset is split into test and train data with the proportion of $0.4$ and $0.6$, respectively.
	
	To better understand the effects of wireless communications channels over the received signals, first, amplitude distributions of four different recordings of all three signals are given in \FGR{fig:measurement_pdfs}. The figure indicates differing power and amplitude levels. The distribution of the received power changes considerably since the measurements are taken at different locations, times and frequency bands. Second, \FGR{fig:measurement_phase_pdfs} illustrates the phase distribution of four different recordings of all three signals. It is seen that the phase of received signals are distributed almost uniformly in between $-\pi$ and $\pi$ radians. This result implies Rayleigh-like fading behavior stemming from the amplitude and phase distributions of received signals. This is an expected result when the measurement area and the locations of transmitters and receivers are considered. Eventually the received power and phase of the signals are obviously affected by the shadowing, multipath fading and path loss as depicted in \FGR{fig:measurement_pdfs} and \FGR{fig:measurement_phase_pdfs}.
	
	The dataset is shared in \cite{safrgh5920} in the format of \ac{SCF}. The dataset covers $60000$ \ac{SCF} matrices with the dimensions of $8193\times16$ corresponding to received \ac{I/Q} samples of $16384$ for each signal.
	
	%%%%%%%%%%%%%%%%%%%%%%%%%%%%%%%%%%%%%%%%%%%%% Measurement PDFs %%%%%%%%%%%%%%%%%%%%%%%
	\begin{figure}[!t]
		\centering
		\includegraphics[width = \linewidth]{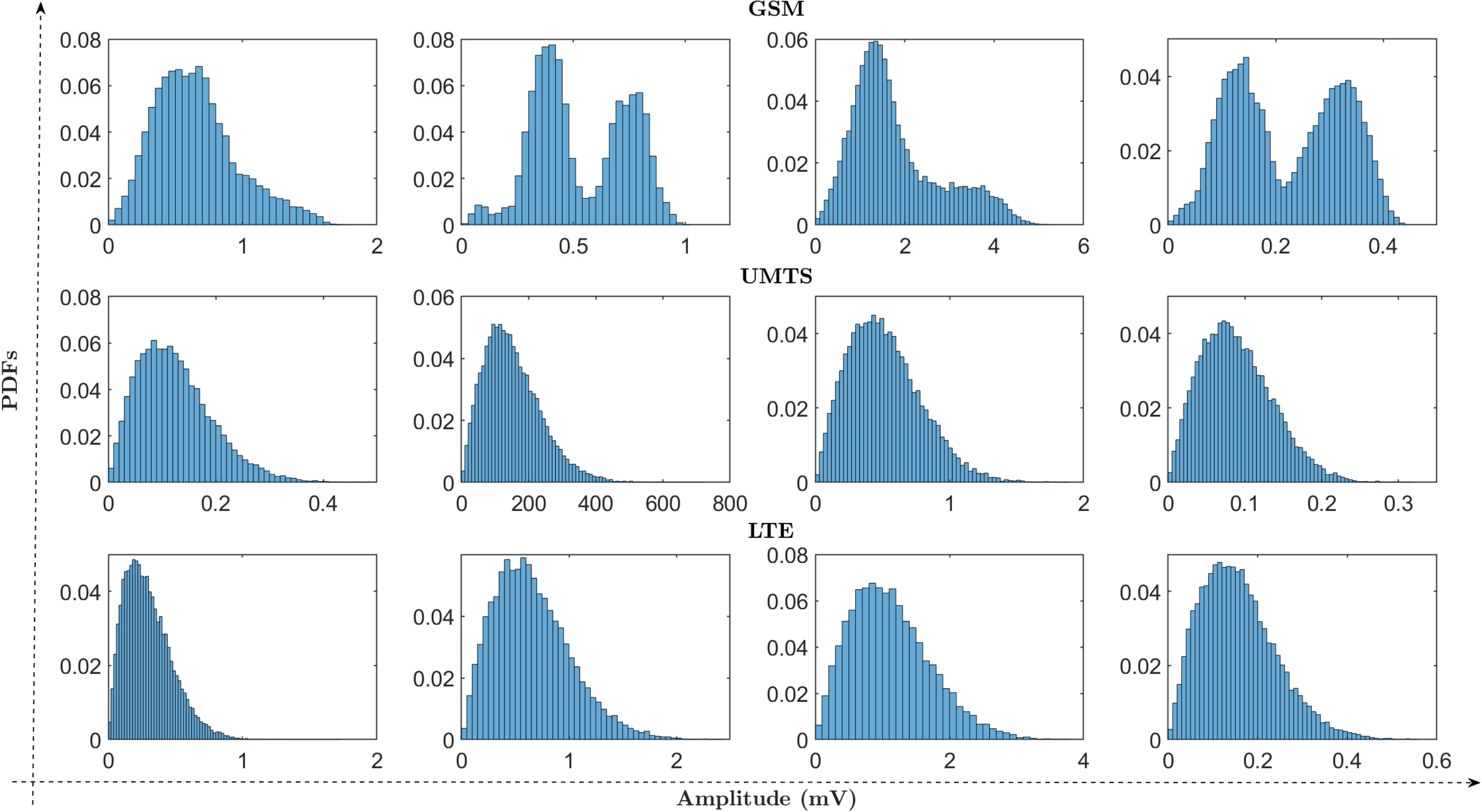}
		\caption{Sample PDFs of the amplitude of received signals in the dataset. The example PDFs show the different channel and received power characteristics.}
		\label{fig:measurement_pdfs}
	\end{figure}

	\begin{figure}[!t]
		\centering
		\includegraphics[width = \linewidth]{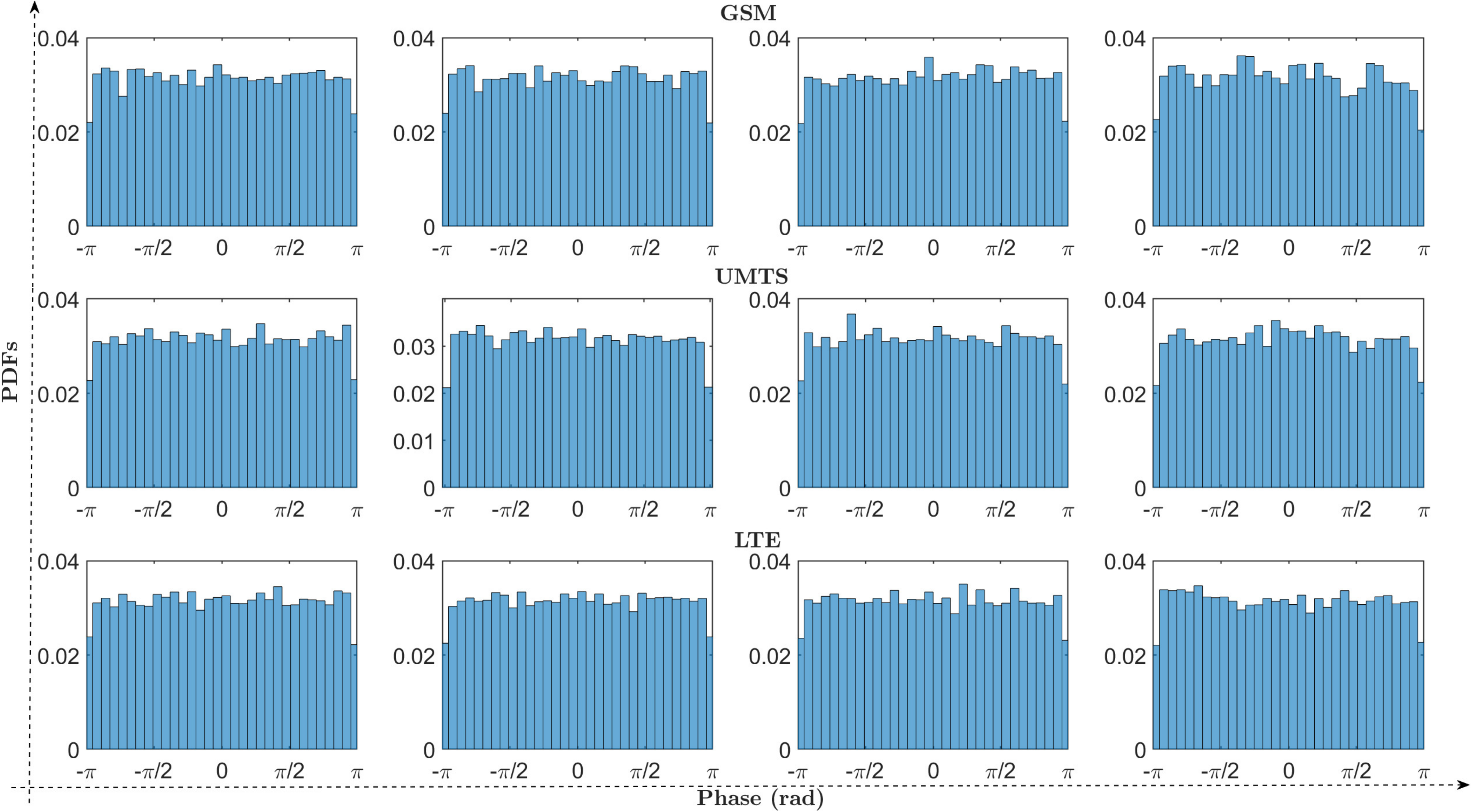}
		\caption{Sample PDFs of the phase of received signals in the dataset. The example PDFs show uniform distribution characteristics.}
		\label{fig:measurement_phase_pdfs}
	\end{figure}
	%%%%%%%%%%%%%%%%%%%%%%%%%%%%%%%%%%%%%%%%%%%%% Measurement PDFs %%%%%%%%%%%%%%%%%%%%%%%

	\section{Classification Performance Analysis}\label{sec:performance}

	\begin{figure}[!t]
		\centering
		\includegraphics[width=\linewidth]{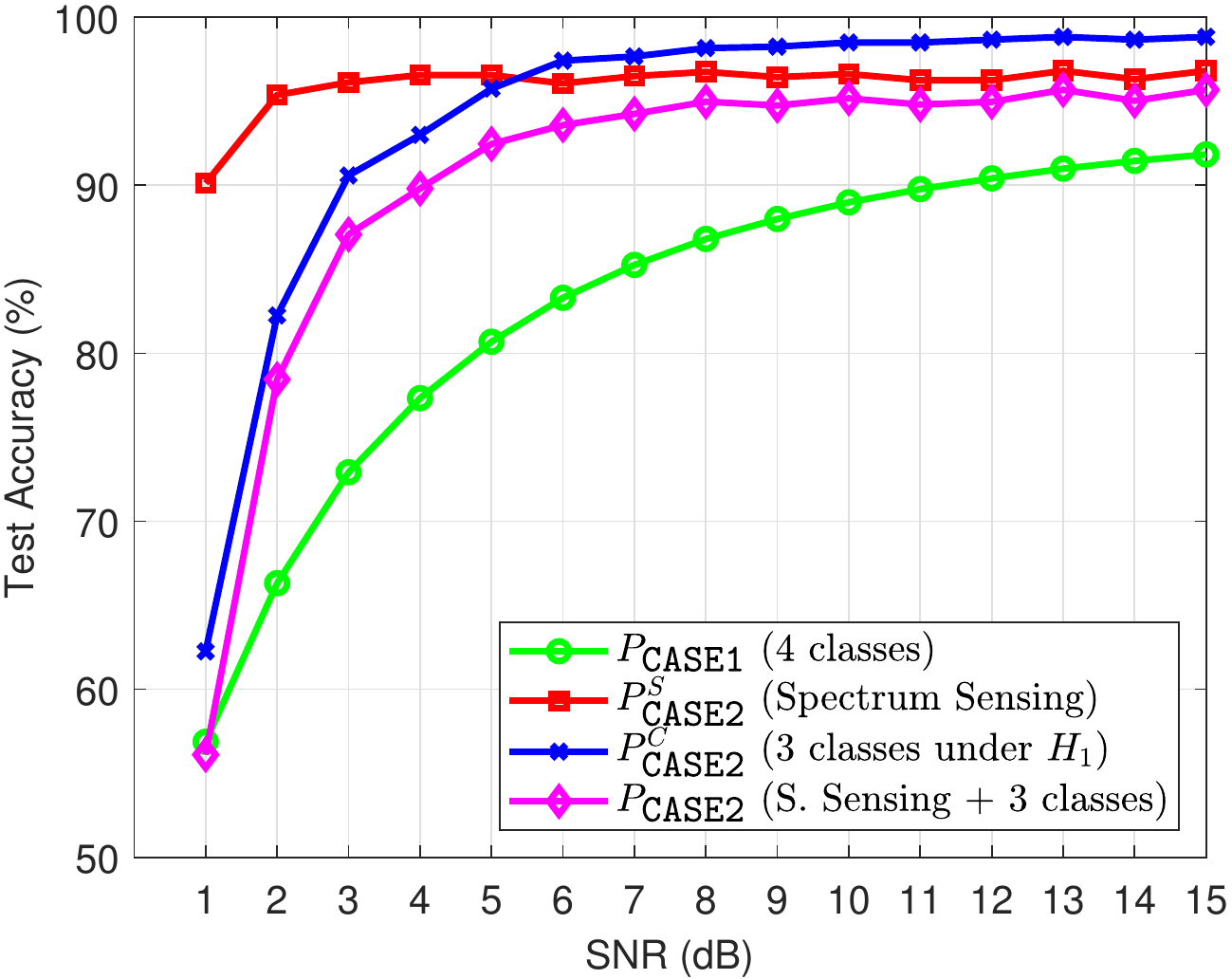}
		\caption{Accuracy values with respect to \ac{SNR} level of the received signals for both cases.}
		\label{fig:acc_all}
	\end{figure}
	
	%%%%%%%%%%%%%%%%%%%%%%%%%%%%%%%%%%%%%%%%%%%%%%%%%%%%%%%%%%%%%%%% TABLE START
	%{\renewcommand{\arraystretch}{0.9}
	\begin{table}[!t]
		\centering
		\caption{Classification performance metrics for the proposed \ac{CNN} model with \ac{SCF}, \ac{AP}, and \ac{FFT} features for \texttt{CASE2}.
		}
		\resizebox{0.90\columnwidth}{!}{
			\begin{tabular}{c c l c c c}
				\toprule
				SNR                                   & Feature              & Signal  & Precision ($\Pi$) & Recall ($\Psi$) & F1-Score \\
				\midrule
				\multicolumn{1}{c}{\multirow{16}{*}{1dB}} & \multirow{4}{*}{I/Q}  & UMTS    & 0.33 & 1.00 & 0.49 \\
				\multicolumn{1}{c}{}                      &                      & LTE     & 0.00 & 0.00 & 0.00 \\
				\multicolumn{1}{c}{}                      &                      & GSM     & 0.00 & 0.00 & 0.00 \\
				\multicolumn{1}{c}{}                      &                      & Average & 0.11 & 0.33 & 0.16 \\
				\cmidrule{2-6}
				\multicolumn{1}{c}{}                      & \multirow{4}{*}{AP}  & UMTS    & 0.35 & 0.39 & 0.37 \\
				\multicolumn{1}{c}{}                      &                      & LTE     & 0.31 & 0.29 & 0.30 \\
				\multicolumn{1}{c}{}                      &                      & GSM     & 0.30 & 0.29 & 0.30 \\
				\multicolumn{1}{c}{}                      &                      & Average & 0.32 & 0.32 & 0.32 \\
				\cmidrule{2-6}
				\multicolumn{1}{c}{}                      & \multirow{4}{*}{FFT} & UMTS    & 0.00 & 0.00 & 0.00 \\
				\multicolumn{1}{c}{}                      &                      & LTE     & 0.21 & 0.50 & 0.30 \\
				\multicolumn{1}{c}{}                      &                      & GSM     & 0.00 & 0.00 & 0.00 \\
				\multicolumn{1}{c}{}                      &                      & Average & 0.07 & 0.17 & 0.10 \\
				\cmidrule{2-6}
				\multicolumn{1}{c}{}                      & \multirow{4}{*}{SCF} & UMTS    & 0.59 & 0.40 & 0.48 \\
				\multicolumn{1}{c}{}                      &                      & LTE     & 0.68 & 0.46 & 0.54 \\
				\multicolumn{1}{c}{}                      &                      & GSM     & 0.62 & 0.98 & 0.76 \\
				\multicolumn{1}{c}{}                      &                      & Average & 0.63 & 0.61 & 0.59 \\
				\midrule
				
				\multirow{16}{*}{5dB}                     & \multirow{4}{*}{I/Q}  & UMTS    & 0.33 & 1.00 & 0.49 \\
				&                      & LTE     & 0.00 & 0.00 & 0.00 \\
				&                      & GSM     & 0.00 & 0.00 & 0.00 \\
				&                      & Average & 0.11 & 0.33 & 0.16 \\
				\cmidrule{2-6}
				& \multirow{4}{*}{AP}  & UMTS    & 0.43 & 0.42 & 0.42 \\
				&                      & LTE     & 0.39 & 0.43 & 0.41 \\
				&                      & GSM     & 0.61 & 0.56 & 0.58 \\
				&                      & Average & 0.47 & 0.47 & 0.47 \\
				\cmidrule{2-6}                                          
				& \multirow{4}{*}{FFT} & UMTS    & 0.53 & 0.49 & 0.51 \\
				&                      & LTE     & 0.25 & 0.51 & 0.34 \\
				&                      & GSM     & 0.00 & 0.00 & 0.00 \\
				&                      & Average & 0.26 & 0.33 & 0.28 \\
				\cmidrule{2-6}
				& \multirow{4}{*}{SCF} & UMTS    & 0.97 & 0.92 & 0.94 \\
				&                      & LTE     & 0.92 & 0.95 & 0.94 \\
				&                      & GSM     & 0.98 & 1.00 & 0.99 \\
				&                      & Average & 0.96 & 0.96 & 0.96 \\
				\midrule
				\multirow{16}{*}{10dB}                     & \multirow{4}{*}{I/Q}  & UMTS    & 0.33 & 1.00 & 0.49 \\
				&                      & LTE     & 0.00 & 0.00 & 0.00 \\
				&                      & GSM     & 0.00 & 0.00 & 0.00 \\
				&                      & Average & 0.11 & 0.33 & 0.16 \\
				\cmidrule{2-6}
				& \multirow{4}{*}{AP}  & UMTS   & 0.50 & 0.52 & 0.51 \\
				&                      & LTE     & 0.50 & 0.53 & 0.52 \\
				&                      & GSM     & 0.88 & 0.79 & 0.83 \\
				&                      & Average & 0.63 & 0.61 & 0.62 \\
				\cmidrule{2-6}                                          
				& \multirow{4}{*}{FFT} & UMTS    & 0.49 & 0.37 & 0.42 \\
				&                      & LTE     & 0.27 & 0.62 & 0.38 \\
				&                      & GSM     & 0.00 & 0.00 & 0.00 \\
				&                      & Average & 0.26 & 0.33 & 0.27 \\
				\cmidrule{2-6}
				& \multirow{4}{*}{SCF} & UMTS    & 1.00 & 0.96 & 0.98 \\
				&                      & LTE     & 0.96 & 0.99 & 0.98 \\
				&                      & GSM     & 1.00 & 1.00 & 1.00 \\
				&                      & Average & 0.99 & 0.98 & 0.98 \\
				\midrule
				\multirow{16}{*}{15dB}                       & \multirow{4}{*}{I/Q}  & UMTS    & 0.33 & 1.00 & 0.49 \\
				&                      & LTE     & 0.00 & 0.00 & 0.00 \\
				&                      & GSM     & 0.00 & 0.00 & 0.00 \\
				&                      & Average & 0.11 & 0.33 & 0.16 \\
				\cmidrule{2-6}
				& \multirow{4}{*}{AP}  & UMTS   & 0.55 & 0.54 & 0.55 \\
				&                      & LTE     & 0.55 & 0.57 & 0.56 \\
				&                      & GSM     & 0.94 & 0.93 & 0.94 \\
				&                      & Average & 0.68 & 0.68 & 0.68 \\
				\cmidrule{2-6}                                          
				& \multirow{4}{*}{FFT} & UMTS    & 0.73 & 0.49 & 0.58 \\
				&                      & LTE     & 0.35 & 0.82 & 0.49 \\
				&                      & GSM     & 0.00 & 0.00 & 0.00 \\
				&                      & Average & 0.36 & 0.44 & 0.36 \\
				\cmidrule{2-6}
				& \multirow{4}{*}{SCF} & UMTS    & 1.00 & 0.97 & 0.99 \\
				&                      & LTE     & 0.97 & 0.99 & 0.98 \\
				&                      & GSM     & 0.99 & 1.00 & 1.00 \\
				&                      & Average & 0.99 & 0.99 & 0.99 \\
				\bottomrule
			\end{tabular} 
		}
		\label{tab:feature_performance}
	\end{table}
	%}
	%%%%%%%%%%%%%%%%%%%%%%%%%%%%%%%%%%%%%%%%%%%%%%%%%%%%%%%%%%%%%% TABLE END

	We evaluate the performance of the proposed classification model over the comprehensive dataset described in \SEC{sec:dataset_generation}. Therefore\textcolor{black}{,} the dataset is composed of \ac{GSM}, \ac{WCDMA} for \ac{UMTS} and \ac{LTE} signals which are recorded over-the-air at  different locations with unique conditions in terms of the number of channel taps, and fading, again as noted in \SEC{sec:dataset_generation}. Training and test sets contain $9000$ and $6000$ signals for each waveform. The \ac{I/Q} signal length is 16384. \ac{CNN} is trained and tested on the \ac{GPU} server equipped with four NVIDIA Tesla V100 \acp{GPU}.
	
	First, we focus on the results for \texttt{CASE1}. As stated before, \texttt{CASE1} refers to four-classes classification problem. As shown in \FGR{fig:acc_all}, the test accuracy of the model exceeds $90\%$ at $11$dB \ac{SNR}. It takes a maximum accuracy value of $92\%$ at $15$dB. The confusion matrices related to \texttt{CASE1} are depicted in \FGR{fig:sc1_conf_matx}. Due to the low \ac{SNR} values, the model mostly can not accurately clasify the signals and identifies the signal as Noise. This case can be observed in \FGR{fig:sc1_conf_matx_1dB}. Therefore\textcolor{black}{,} dividing the problem into two parts becomes a viable alternative: first sense, then classify. In this case, we analyse both \ac{CNN} detector and \ac{CNN} classifier (see \FGR{fig:case1_case2}). For the sensing part of the architecture, noise signals are labeled as $0$ and the rest of the set is labeled as $1$. The detection results are plotted again in \FGR{fig:acc_all} as $P^{\texttt{S}}_{\texttt{CASE2}}$. The detection accuracy follows $96\%$ at almost all \ac{SNR} values. 
	
	Following the steps above, assuming that the signal is present in the spectrum at the output of CNN detector of \texttt{CASE2} in \FGR{fig:case1_case2}, the performance of the \ac{CNN} classifier can be investigated. This stage is labeled as  $P^{\texttt{C}}_{\texttt{CASE2}}$ in \FGR{fig:acc_all} and it is observed that the classification accuracy exceeds $90\%$ at $3$dB \ac{SNR}. It gives the best performance, $98.5\%$, at $9$dB and it is remained  stable until $15$dB. The  \FGR{fig:sc2_conf_matx_1dB} depicts the confusion matrices related to CNN classifier of \texttt{CASE2} and implies that even at low \ac{SNR} regime, the classifier can identify \ac{GSM} signals with high accuracy; however, overall precision of the classifier is low \textit{i.e.},  in contrary to \ac{GSM} signals, the classifier has difficulty in recognition of \ac{UMTS} and \ac{LTE} signals in low SNR regime. But the accuracy and precision of the classifier enhance as \ac{SNR} increases  in \FGR{fig:sc2_conf_matx_5dB} and \FGR{fig:sc2_conf_matx_10dB}. \textcolor{black}{This phenomena is observed due to the dominance of characteristics in feature matrices which follow Gaussian distribution. As known, GSM is associated with Gaussian minimum shift keying (GMSK); therefore, GSM signals inherently show characteristics defined by Gaussian distribution in the case of high SNR. Decreasing in SNR leverages Gaussian characteristics in the received signal because of \ac{AWGN}. That is to say, UMTS and LTE signals with lower SNR values become denoting Gaussian characteristics; thus, the model is prone to learn Gaussian characteristics to decrease its loss function. When the trained model is tested, it is expected that the model can identify the signals which have dominant Gaussian characteristics. As a result, the model can identify GSM signals, which inherently denote Gaussian attributes, in lower SNR regime where UMTS and LTE signals lose their unique features. This statement shows parallelism with the results given in \texttt{CASE1} in~\FGR{fig:sc1_conf_matx_1dB}. The model accurately identifies \ac{AWGN} at lower SNR regime as given in~\FGR{fig:sc1_conf_matx_1dB}.}
	%%%%%%%%%%%%%%%%%%%%%%%%%%%%% Confusion Matrices : Case 1 %%%%%%%%%%%%%%%%%%%%%%%
	\begin{figure*}[!t]
		\centering
		\subfigure[]{
			\label{fig:sc1_conf_matx_1dB}
			\includegraphics[width=0.32\linewidth]{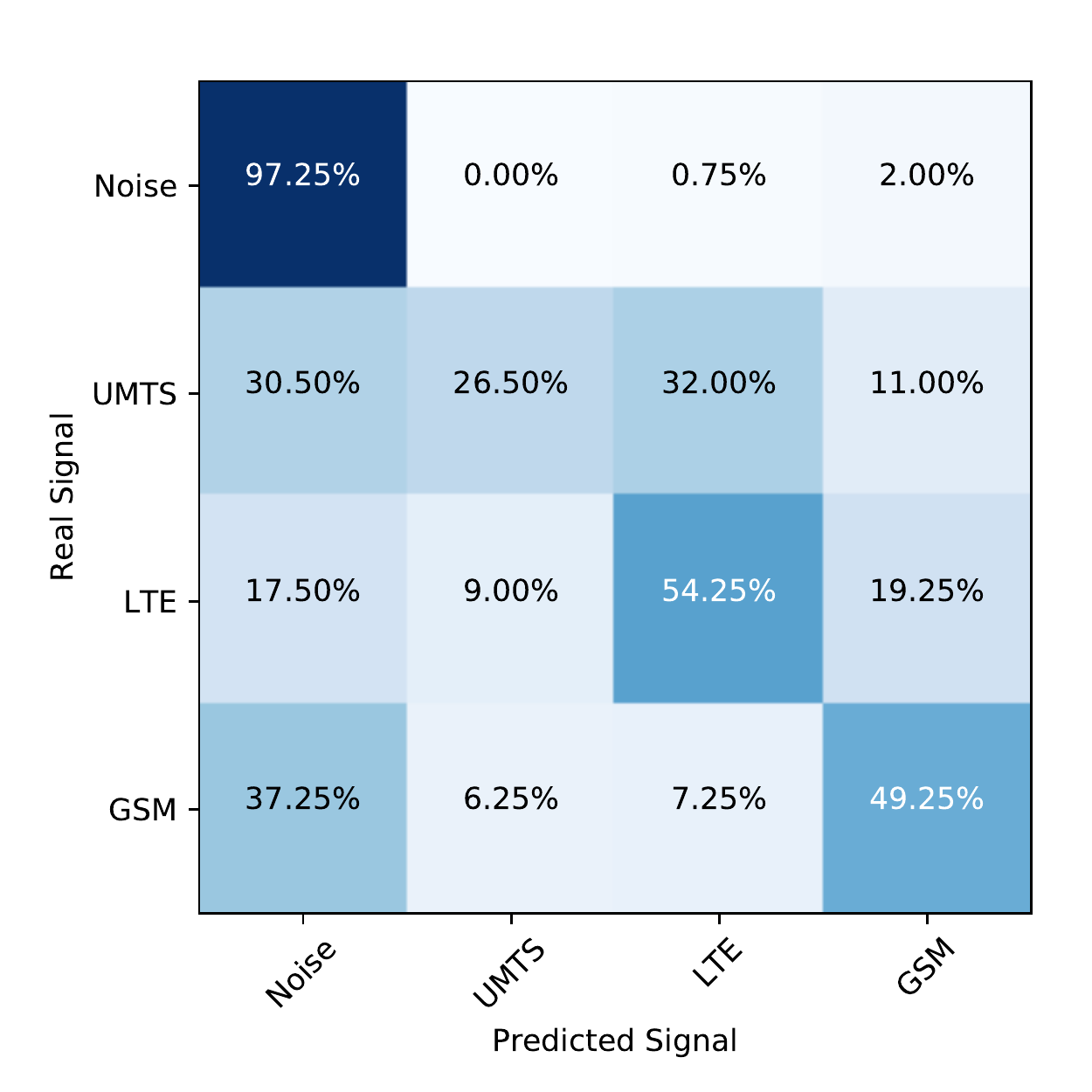}}
		\subfigure[]{
			\label{fig:sc1_conf_matx_5dB}
			\includegraphics[width=0.32\linewidth]{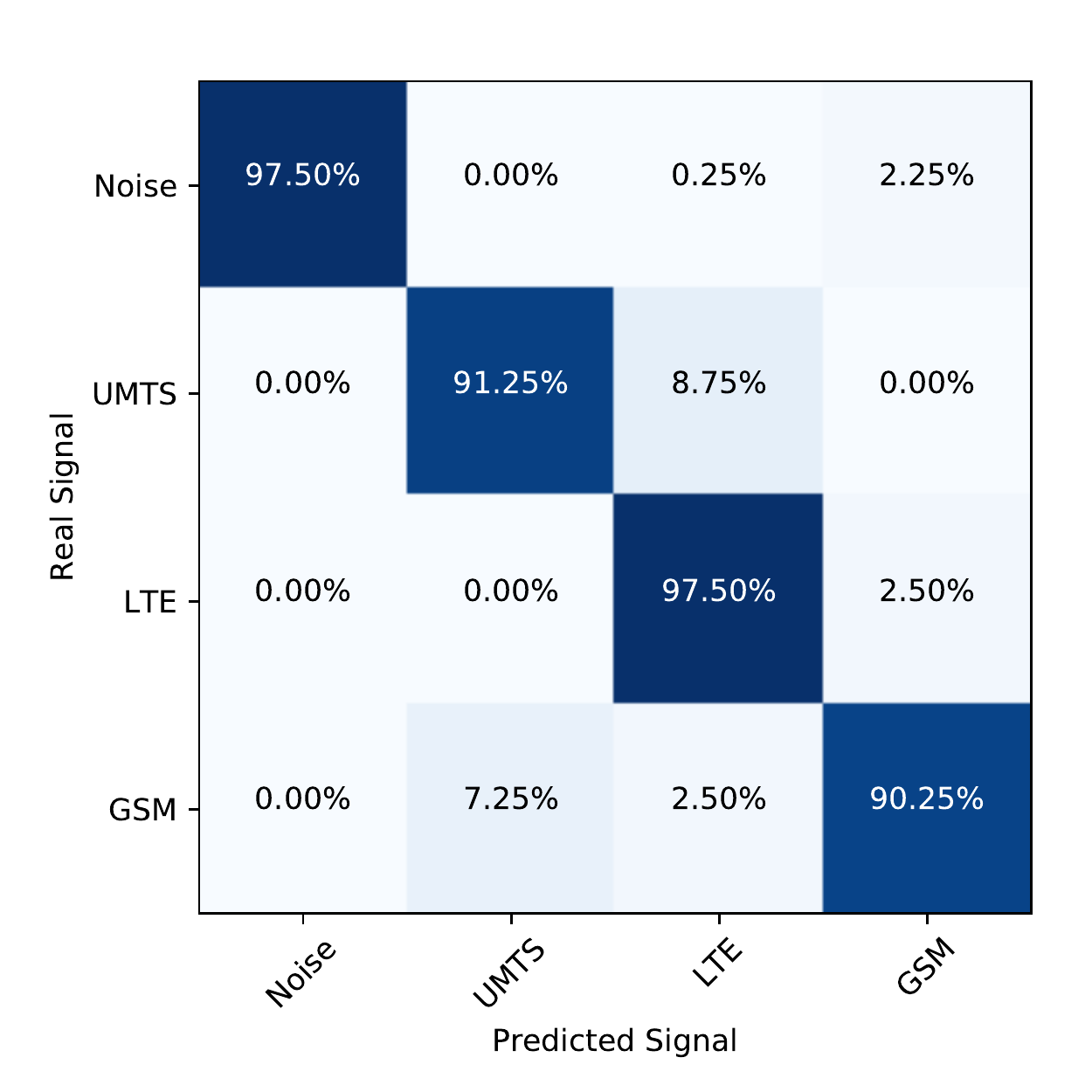}}
		\subfigure[]{
			\label{fig:sc1_conf_matx_10dB}
			\includegraphics[width=0.32\linewidth]{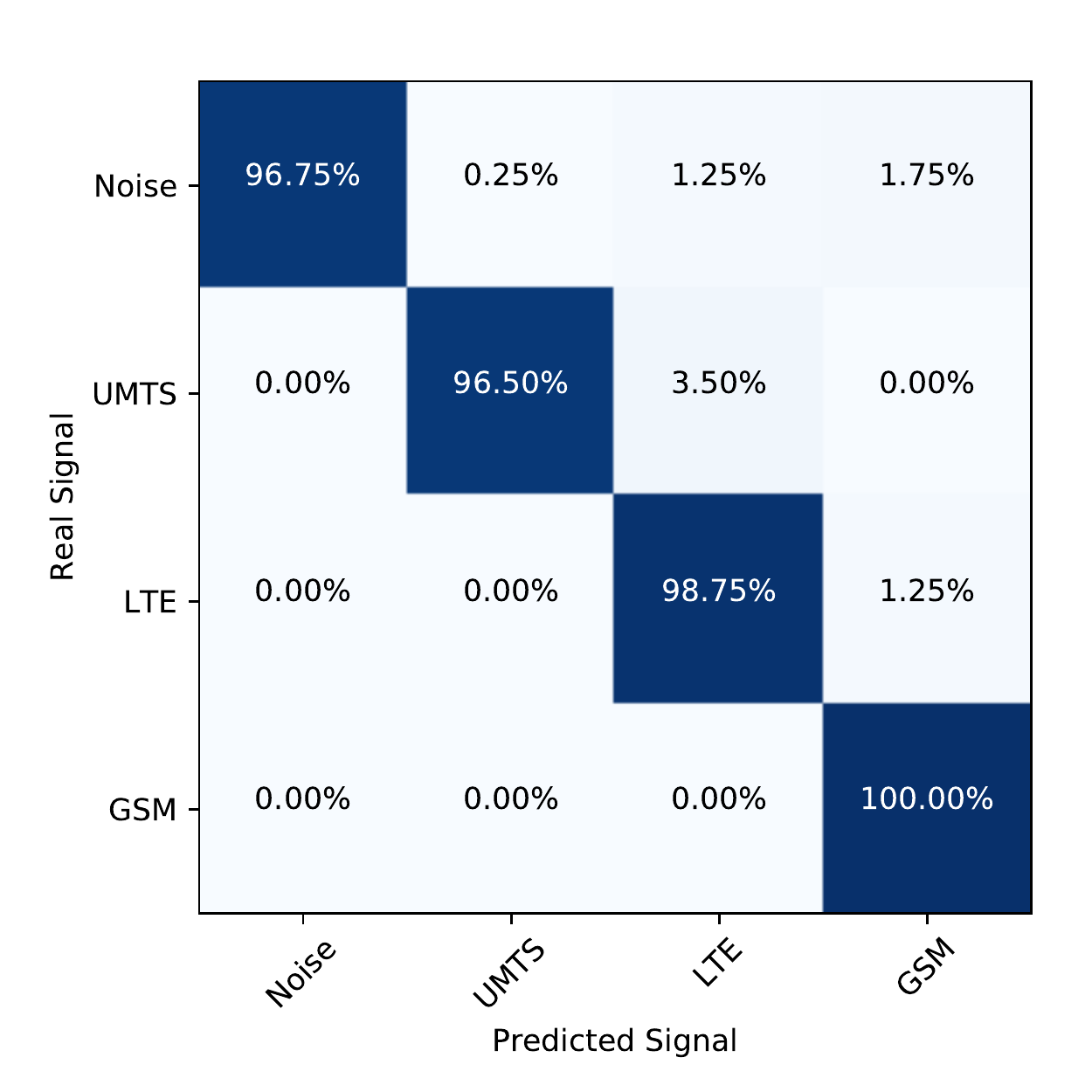}}
		\caption{Confusion matrices for \texttt{CASE1} at \ac{SNR} levels of (a) 1dB, (b) 5dB, and (c) 10dB. It should be noticed that the model does not randomly choose only one signal at low SNR level.}
		\label{fig:sc1_conf_matx}
	\end{figure*}
	%%%%%%%%%%%%%%%%%%%%%%%%%%%%% Confusion Matrices : Case 1 %%%%%%%%%%%%%%%%%%%%%%% %%%%%%%%%%%%%%%%%%%%%%%%%%%%% Confusion Matrices : Case 2 %%%%%%%%%%%%%%%%%%%%%%%
	\begin{figure*}[!t]
		\centering
		\subfigure[]{
			\label{fig:sc2_conf_matx_1dB}
			\includegraphics[width=0.32\linewidth]{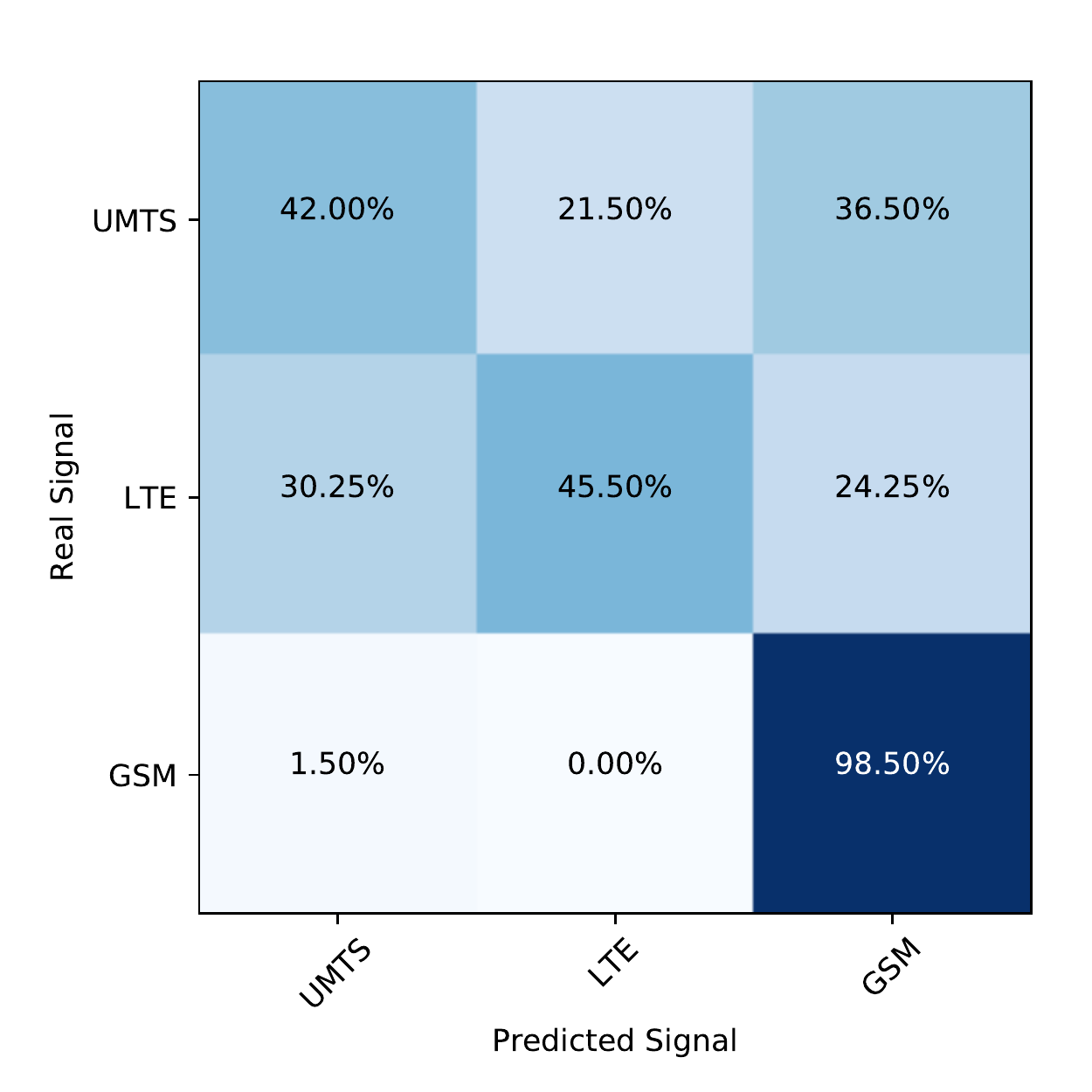}}
		\subfigure[]{
			\label{fig:sc2_conf_matx_5dB}
			\includegraphics[width=0.32\linewidth]{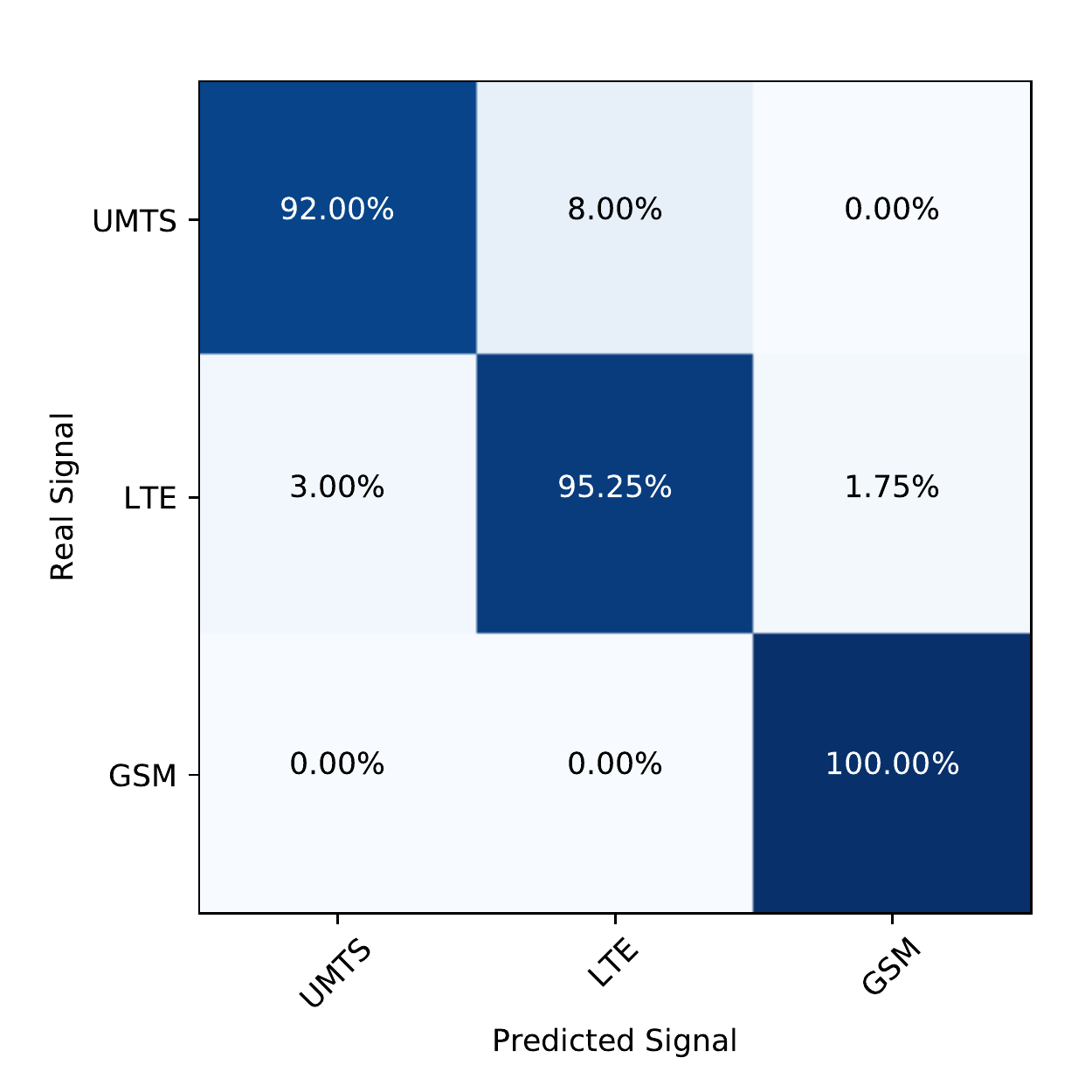}}
		\subfigure[]{
			\label{fig:sc2_conf_matx_10dB}
			\includegraphics[width=0.32\linewidth]{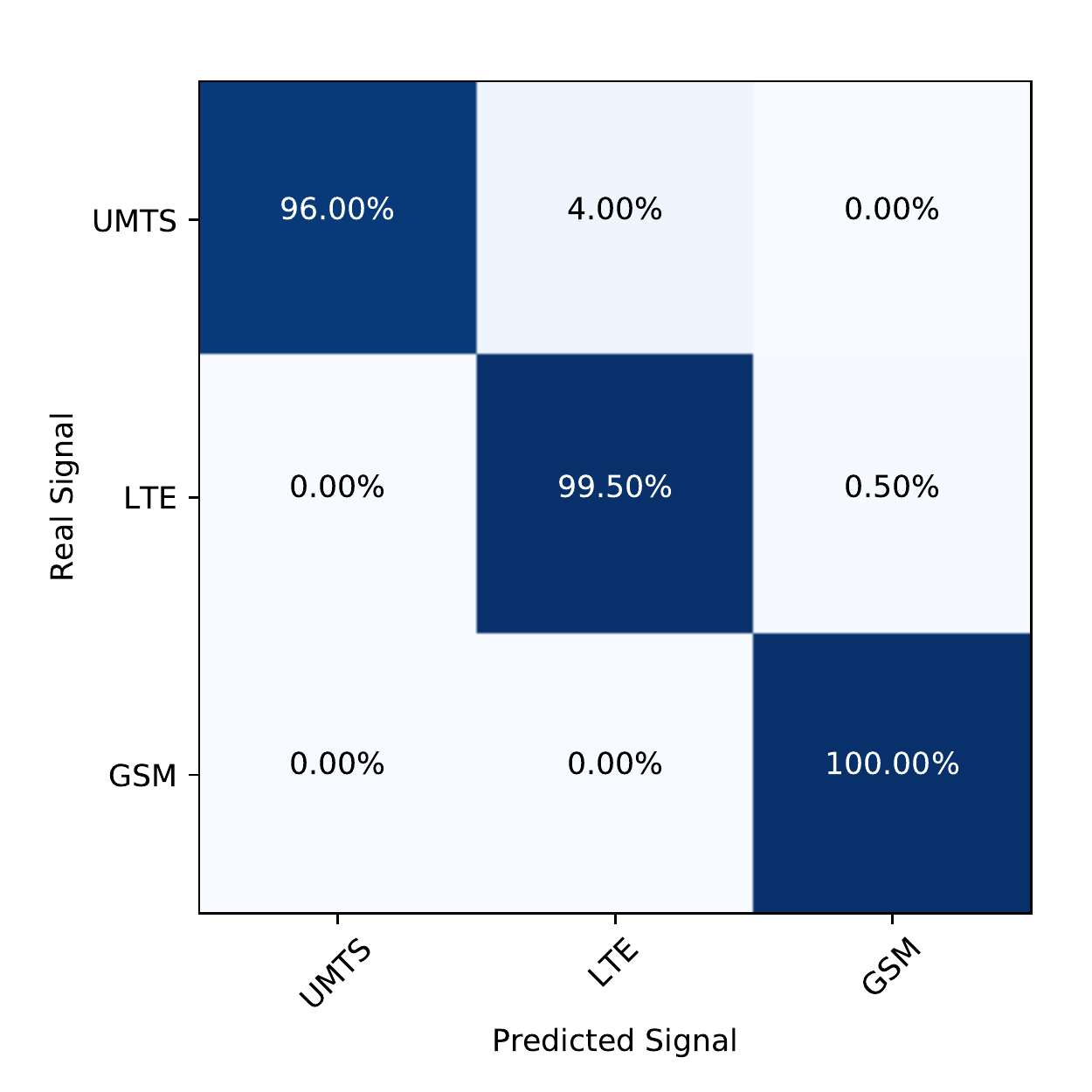}}
		\caption{Confusion matrices for the classification part of \texttt{CASE2} at \ac{SNR} levels of (a) 1dB, (b) 5dB, and (c) 10dB. It should be noticed that the model does not randomly choose only one signal at low SNR level.}
		\label{fig:sc2_conf_matx}
	\end{figure*}
	%%%%%%%%%%%%%%%%%%%%%%%%%%%%% Confusion Matrices : Case 2 %%%%%%%%%%%%%%%%%%%%%%%
	
	The results for \texttt{CASE2} are given in parts to this point. Now, we can examine the overall performance of \texttt{CASE2}. Obviously, there is a loss of performance due to some misdetection in the sensing phase. Both the detection rate in the sensing stage and the accuracy in the classification stage are high at $3$dB and thereafter, so overall performance does not suffer a significant loss. As shown in \FGR{fig:acc_all}, the overall performance of \texttt{CASE2} is far superior to that of \texttt{CASE1}. Especially at low \ac{SNR} levels, the signals remaining after first detecting and separating noise from the signal set by the \ac{CNN} detector can be classified with much better performance. In this way, the performance is higher in \texttt{CASE2}. However, it should be noted that \texttt{CASE2} is more costly than \texttt{CASE1} in terms of training time and the number of models. Obviously, \texttt{CASE2} can be predicted to perform better than \texttt{CASE1} in the presence of a jammers exhibiting Gaussian characteristics or other interfering signals.
	
	\textcolor{black}{Sensing performance can be considered that it is slightly lower than conventional spectrum sensing methods like energy detector and matched filter. However, it should be noted that this study employs real--world data rather than simulation or synthetic data. For example, energy detectors can sense a signal in a spectrum with optimal performance; however, it needs to know noise variance. But even with a slight error on estimating the noise variance, the sensing performance seriously decreases.  Moreover, as the power of spread spectrum signals (e.g. WCDMA in UMTS) is spreaded in a wide band, its power is very close to the noise floor. By taking into this account, in a fading environment, it can be said that ED cannot perform a satisfactory detection rate for spread spectrum signals as stated in~\cite{cabric2004implementation}. It is worth noting that our measurements follow Rayleigh distribution as seen in \FGR{fig:measurement_pdfs} and \FGR{fig:measurement_phase_pdfs}. On the other hand, matched filters are waveform-specific solution and they require the perfect knowledge for signals.}
	
	\subsection{Investigation for the Impact of Different Features}
	
	In this section\textcolor{black}{,} we compare the performance of other features of \ac{I/Q}, \ac{AP}, and \ac{FFT} which are frequently employed for sensing purposes with \ac{SCF}. The features are used as detailed in \SEC{sec:background}. The results of this test are presented in \TAB{tab:feature_performance}. Unlike the modulation classification studies~\cite{o2018over, tekbiyikvtc2020}, \ac{I/Q} cannot provide a meaningful input for the model due to the severe fading effect on the phase of signal, which is seen in \FGR{fig:measurement_phase_pdfs}. The histograms of phase imply that the signal phase \textcolor{black}{is} corrupted and the information on the phase is lost. That is why \ac{I/Q} shows poor performance. The average performances also indicate that \ac{SCF} outperforms \ac{I/Q}, \ac{AP}, and \ac{FFT} for all \ac{SNR} levels. Assuming that these two are used along with \ac{I/Q} as the main features for training, these results show significant gains for real-world signals especially above \unit{$5$}{dB} \ac{SNR} level. It is observed that \ac{AP} performs better than \ac{FFT}. The average training time per epoch is approximately \unit{$60$}{s} for \ac{SCF} feature where both \ac{FFT} and \ac{AP} take \unit{$7.5$}{s} per epoch; however, both \ac{FFT} and \ac{AP} cannot show an acceptable classification performance, $P_{\texttt{CASE2}}^{C}$. Although the cost of computing both features is far behind the \ac{SCF}, they are far from delivering the desired performance. To visualize the vectors in input space, we employ the \mbox{\acl{t-SNE}} (t-SNE) algorithm. Although originally \mbox{\ac{I/Q}} samples are not linearly separable, \mbox{\ac{SCF}} clusters the vectors in the space and allow\textcolor{black}{s} almost linear separation as depicted in \mbox{\FGR{fig:tsne}}. The analysis based on t-SNE results show that \mbox{\ac{SCF}} better separates signal vectors in space. The results of this study are in line with the previous analysis~\mbox{\cite{tekbiyik2019multi}}.
	
	\begin{figure}[h]
		\centering    
		\subfigure[I/Q]{%
			\label{fig:IQ_feature_tsne}%
			\includegraphics[width=0.19\textwidth]{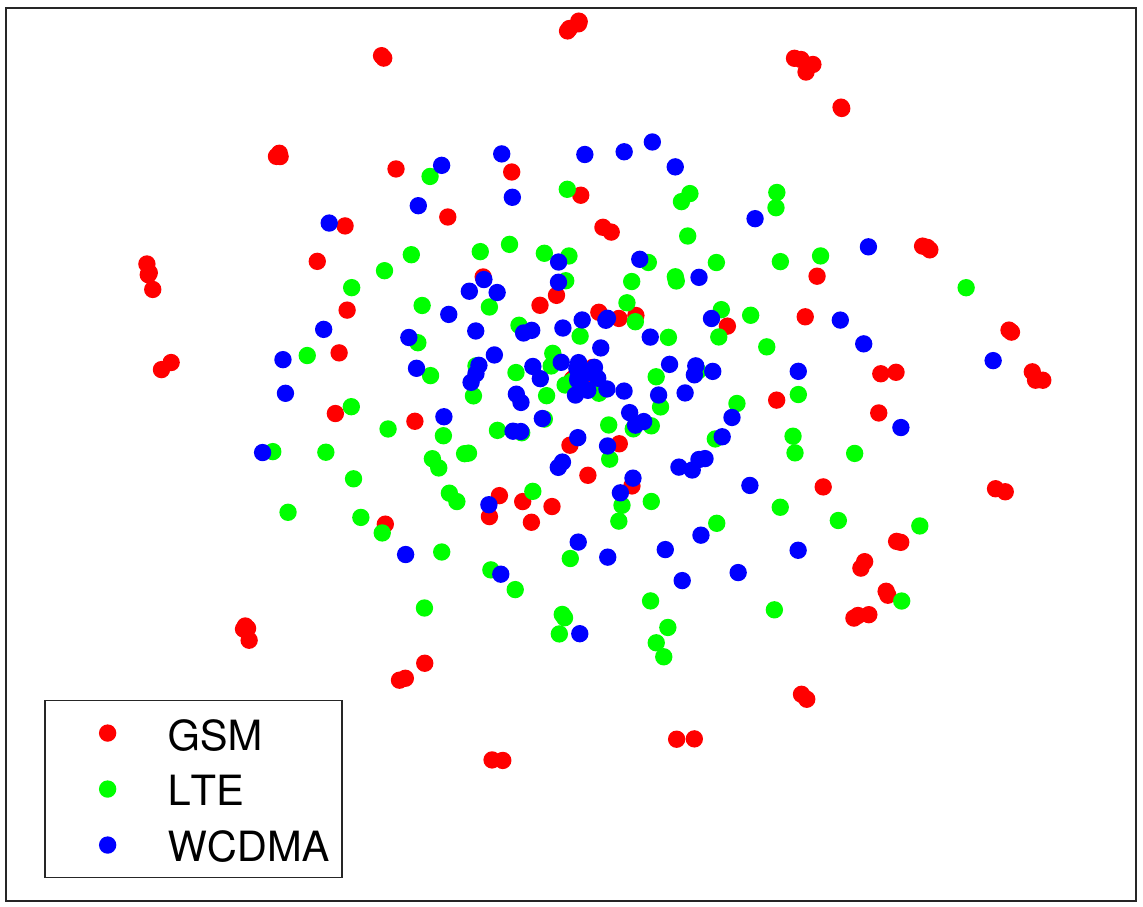}}%
		\quad
		\subfigure[AP]{%
			\label{fig:AP_feature_tsne}%
			\includegraphics[width=0.19\textwidth]{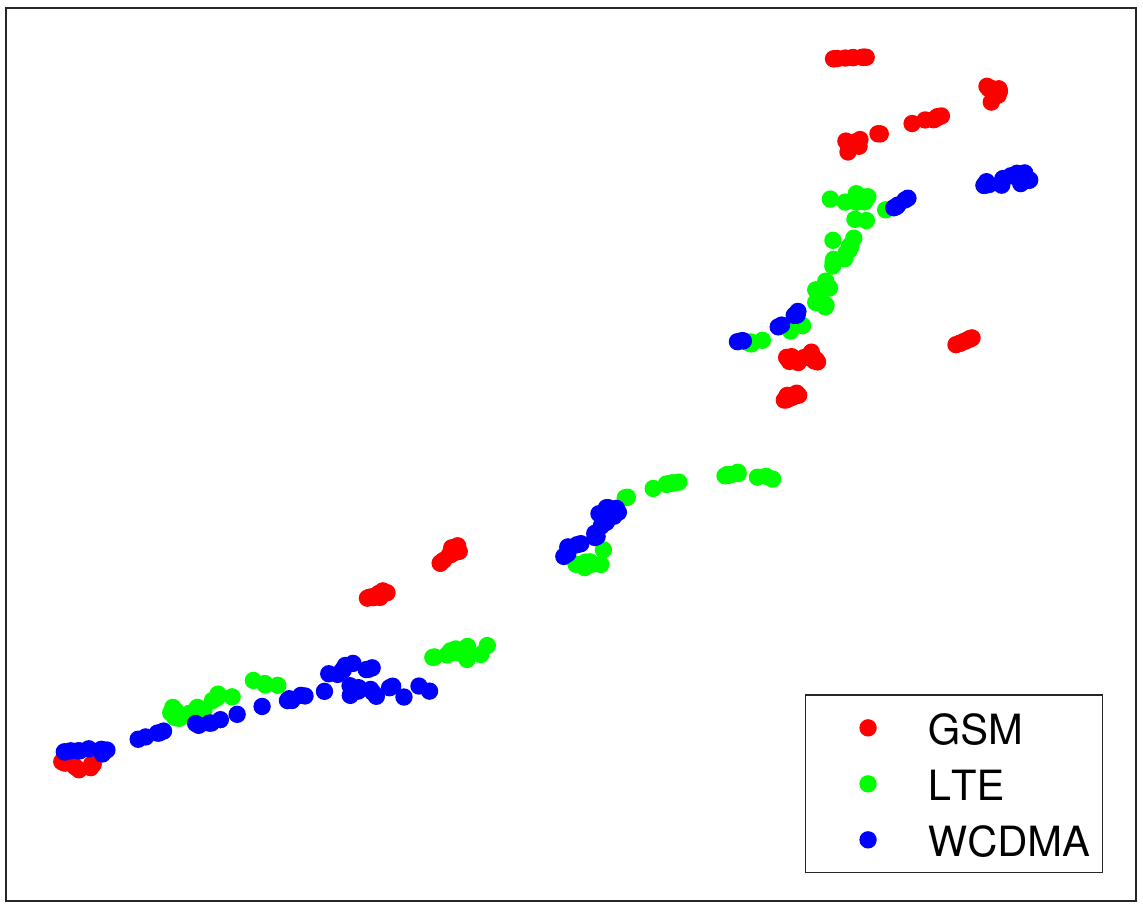}}%
		\quad
		\subfigure[FFT]{%
			\label{fig:FFT_feature_tsne}%
			\includegraphics[width=0.19\textwidth]{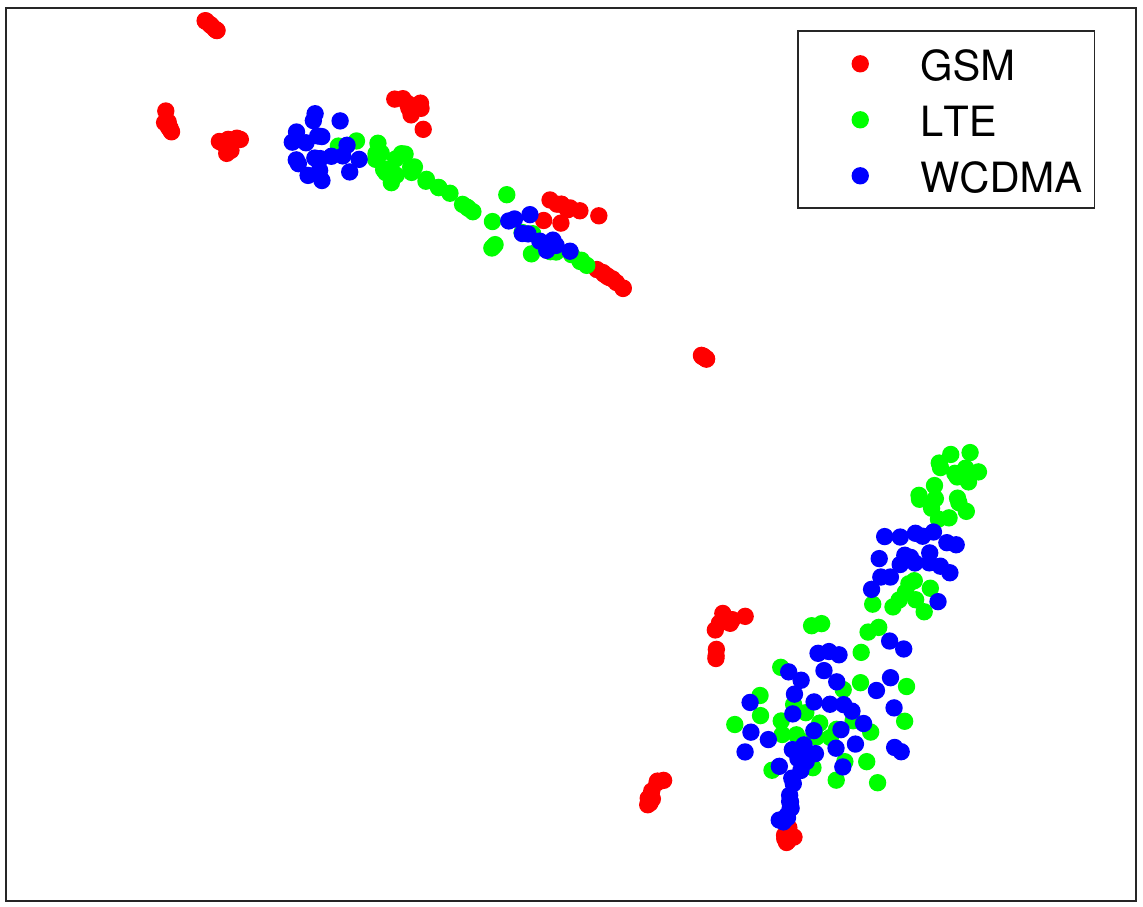}}
		\quad
		\subfigure[SCF]{%
			\label{fig:SCF_feature_tsne}%
			\includegraphics[width=0.19\textwidth]{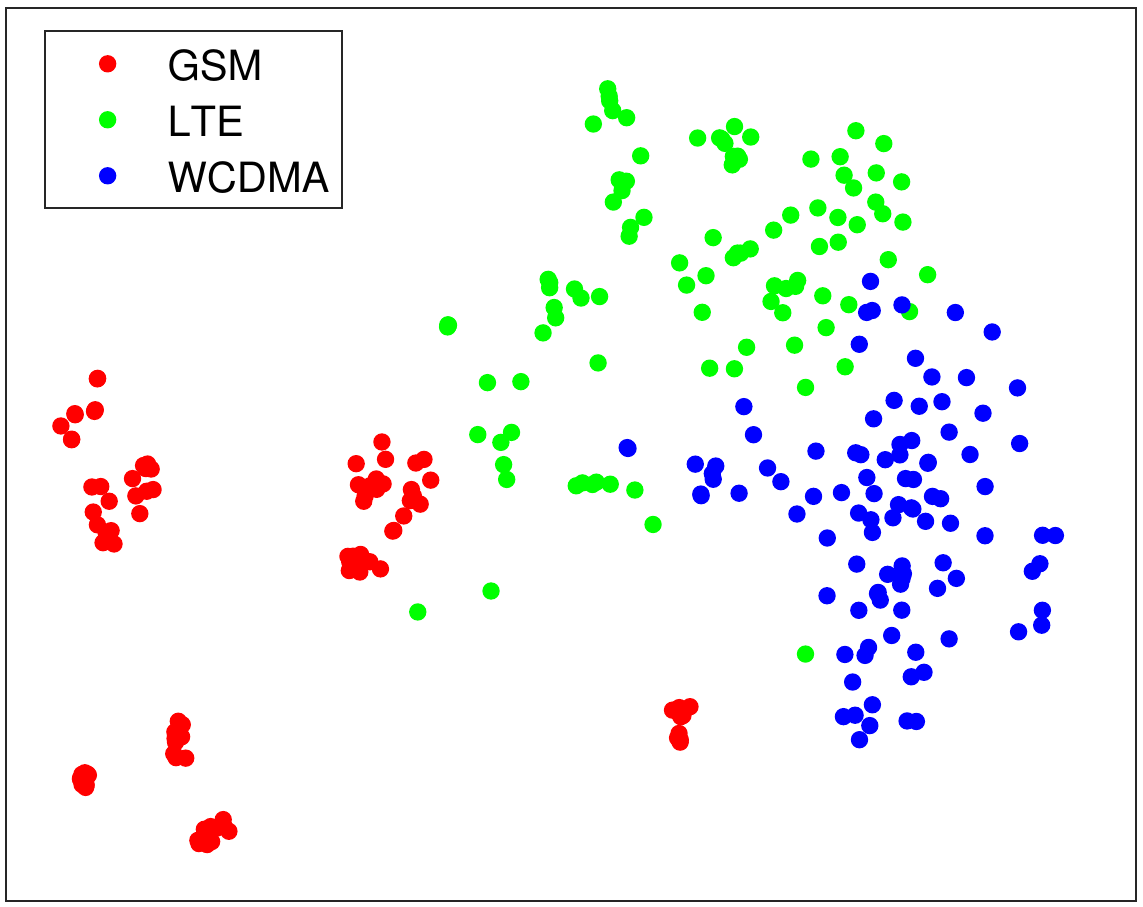}}
		\caption{\hl{Two-dimensional demonstration of the features by the t-SNE algorithm. This illustration shows that in contrary to the other features, SCF can separately cluster real-world signals in space successfully.}}
		\label{fig:tsne}
	\end{figure}
	
	\subsection{Comparison with Existing Deep Learning Networks}
	%%%%%%%%%%%%%%%%%%%%%%%%%%%%%%%%%%%%%%%%%%%%%%%%%%%%%%%%%%%%%%%%%% DL TABLE
	\begin{table}[!t]
		\centering
		\caption{Performance comparison between the existing \ac{DL} networks and the proposed system for the classification stage of {\texttt{CASE2}} at \ac{SNR} value of 15dB.}\label{tab:dl_performance}
		\begin{tabular}{c l c c c }
			\toprule
			Network                                  & Signal  & \textbf{Precision} & \textbf{Recall} & \textbf{$F_1$-score} \\ \midrule%\hline
			\multicolumn{1}{c}{\multirow{4}{*}{CLDNN \cite{ramjee2019fast}}}   & UMTS   &    0.33   &   1.00   &  0.50      \\ %\cline{2-5}  
			\multicolumn{1}{c}{}                                  & LTE    &    0.00   &   0.00    &  0.00      \\ %\cline{2-5}  
			\multicolumn{1}{c}{}                             & GSM   &    0.00   &  0.00    &  0.00      \\ %\cline{2-5} 
			\multicolumn{1}{c}{}                                 & Average &    0.11 &    0.33   &  0.17            \\ \midrule%\hline
			\multirow{4}{*}{LSTM \cite{hu2018robust}}           & UMTS   &    0.33  &   1.00   &   0.50       \\ %\cline{2-5}  
			& LTE    &   0.00    &  0.00   &   0.00       \\ %\cline{2-5}  
			& GSM     &    0.00    &  0.00   &   0.00       \\ %\cline{2-5}  
			& Average &    0.11     & 0.33    &  0.17       \\ \midrule%\hline
			\multirow{4}{*}{Proposed CNN with SCF}                     & UMTS   &    0.79 & 1.00 & 0.88      \\ %\cline{2-5} 
			& LTE     & 1.00 & 0.72 & 0.84      \\ %\cline{2-5}  
			& GSM     & 0.99 & 1.00 & 0.99       \\ %\cline{2-5}  
			& Average &    0.93 & 0.91 & 0.91      \\ 
			\bottomrule%\hline
		\end{tabular}
	\end{table}
	%%%%%%%%%%%%%%%%%%%%%%%%%%%%%%%%%%%%%%%%%%%%%%%%%%%%%%%%%%%%%%%%%% DL TABLE
	The existing \ac{DL} networks are employed to classify the cellular communication signals. We utilize \ac{CLDNN} {\cite{ramjee2019fast}} and \ac{LSTM} {\cite{hu2018robust}} models. These models are originally used in modulation classification. Without any change in the models, input matrix, and input vector as proposed in the papers are adopted in the study. \ac{CLDNN} takes a $2\times128$ matrix which is composed of amplitude and phase values for each \ac{I/Q} sample. On the other hand, \ac{LSTM} model utilizes a vector reshaped version of the matrix used in \ac{CLDNN}. Therefore, the length of the vector is $256$. Its first half includes in-phase components while the rest of the vector is quadrature components. Other details are found in \cite{ramjee2019fast, hu2018robust}. The precision, recall, and $F_{1}$-score are given in \TAB{tab:dl_performance}. It shows that \ac{CLDNN} and \ac{LSTM} decide that the received signal is \ac{UMTS} whatever it actually is. Even though \ac{LSTM} and \ac{CLDNN} can be trained in a short time by using \ac{I/Q} vector and matrix, employing \ac{I/Q} vector and matrix give poor classification performance.
	
	\subsection{Comparison with SVM}
	
	In our previous work\textcolor{black}{,} we employed \ac{SVM}s to identify real-world signals \cite{tekbiyik2019multi}. Even though utilization of \ac{SCF} in \ac{SVM} provides good performance, \hl{training of \mbox{\ac{SVM}} should be conducted for each \mbox{\acs{SNR}} level separately \textit{i.e.}, at the end of the training, the more \mbox{{SNR}} values in the dataset, the more models should be created. The real-world utilization of \mbox{\ac{SVM}} requires an \mbox{\acs{SNR}} estimator and loading of all pre-trained models to memory during operation; thus, reducing the applicability of the method and making improvements a necessity.} As seen in \FGR{fig:acc_svm_vs_cnn_pcase2_c}, the \ac{CNN}-based classifier shows a superior performance compared to \ac{SVM}-based classifier of \cite{tekbiyik2019multi}, under the conditions of the classification part of \texttt{CASE2}. To this end, while \ac{CNN}-based classifier employs a less costly feature due to elimination of mapping of bi-frequency spectrum, it still performs with higher accuracy. Therefore\textcolor{black}{,} producing a model independent of the \mbox{\ac{SNR}} is an advantage of the proposed \mbox{\ac{CNN}} based method since the training set contains an equal number of signals from each \mbox{\ac{SNR}}. As a result, a single model would be adequate for classification in a large \mbox{\ac{SNR}} range at the test stage.
	
	\begin{figure}[!t]
		\centering
		\includegraphics[width=\linewidth]{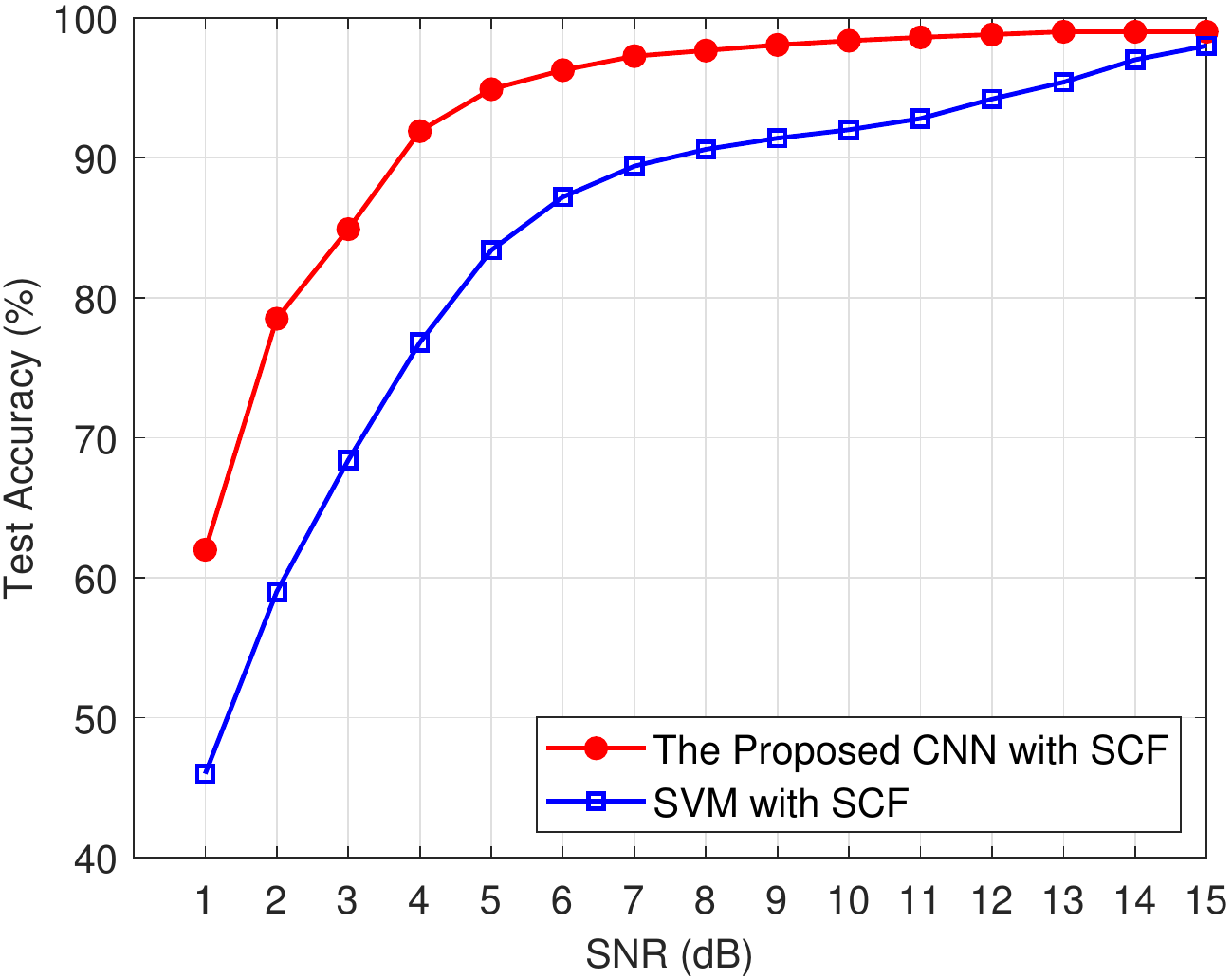}
		\caption{The classification performance comparison between SVM in~\cite{tekbiyik2019multi} and the proposed \ac{CNN} structure for $P_{\texttt{CASE2}}^{\texttt{C}}$.}
		\label{fig:acc_svm_vs_cnn_pcase2_c}
	\end{figure}
	
	\begin{figure}[!t]
		\centering
		\includegraphics[width=\linewidth]{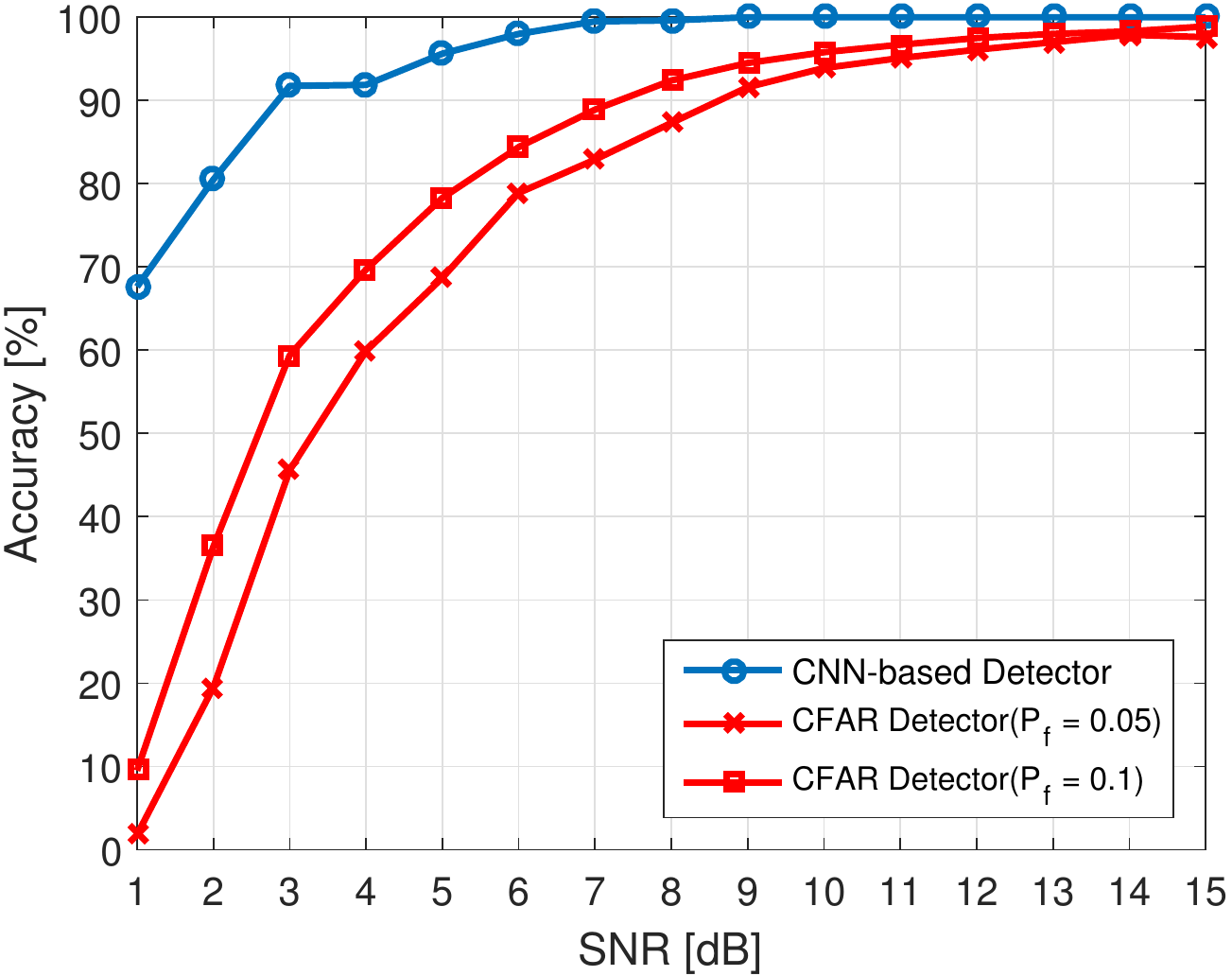}
		\caption{Spectrum sensing performances of CFAR detectors and \ac{CNN}-based detector with respect to \ac{SNR}.}
		\label{fig:spectrum_sensing}
	\end{figure}
	
	\subsection{Comparison with CFD}
	
	Besides signal classification, the proposed \ac{CNN} model can be used for spectrum sensing. We investigated the sensing performance of the model by training a \ac{CNN}-based spectrum occupancy detector trained over 600 pure noise signals and 600 noisy \ac{WCDMA} signals for each \ac{SNR} value. Then, the model is tested with 400 pure noise signals and 400 noisy \ac{WCDMA} signals for each \ac{SNR} level and sensing results are acquired. Furthermore, for comparison purposes, we implement a \textcolor{black}{\ac{CFAR}} detector utilizing classical \ac{CFD} \cite{spooner2018wideband} to identify \ac{WCDMA} signals and the same dataset is also used for \ac{CFAR} detector. Please note that \ac{UMTS} signals are deliberately selected due to their known dominant \ac{SCF} characteristics stemming from cyclic spreading codes. The results of this test are given in \FGR{fig:spectrum_sensing}. In view of these results, it is clearly seen that the \ac{CNN}-based detector outperforms the \ac{CFAR} detector at all \ac{SNR} regimes. For example, the sensing performance of the \ac{CNN}-based detector is $91.75\%$ at \unit{$3$}{dB} while the probability of detection for the \ac{CFAR} detector are $45.6\%$ and $59.4\%$ for the selected false alarm rates as $0.05$ and $0.1$, respectively.
	
	\subsection{Focusing on the Meaningful Region of Spectral Correlation Function}\label{sec:focus}
	
	\begin{figure}[!t]
		\centering
		\includegraphics[width=\linewidth]{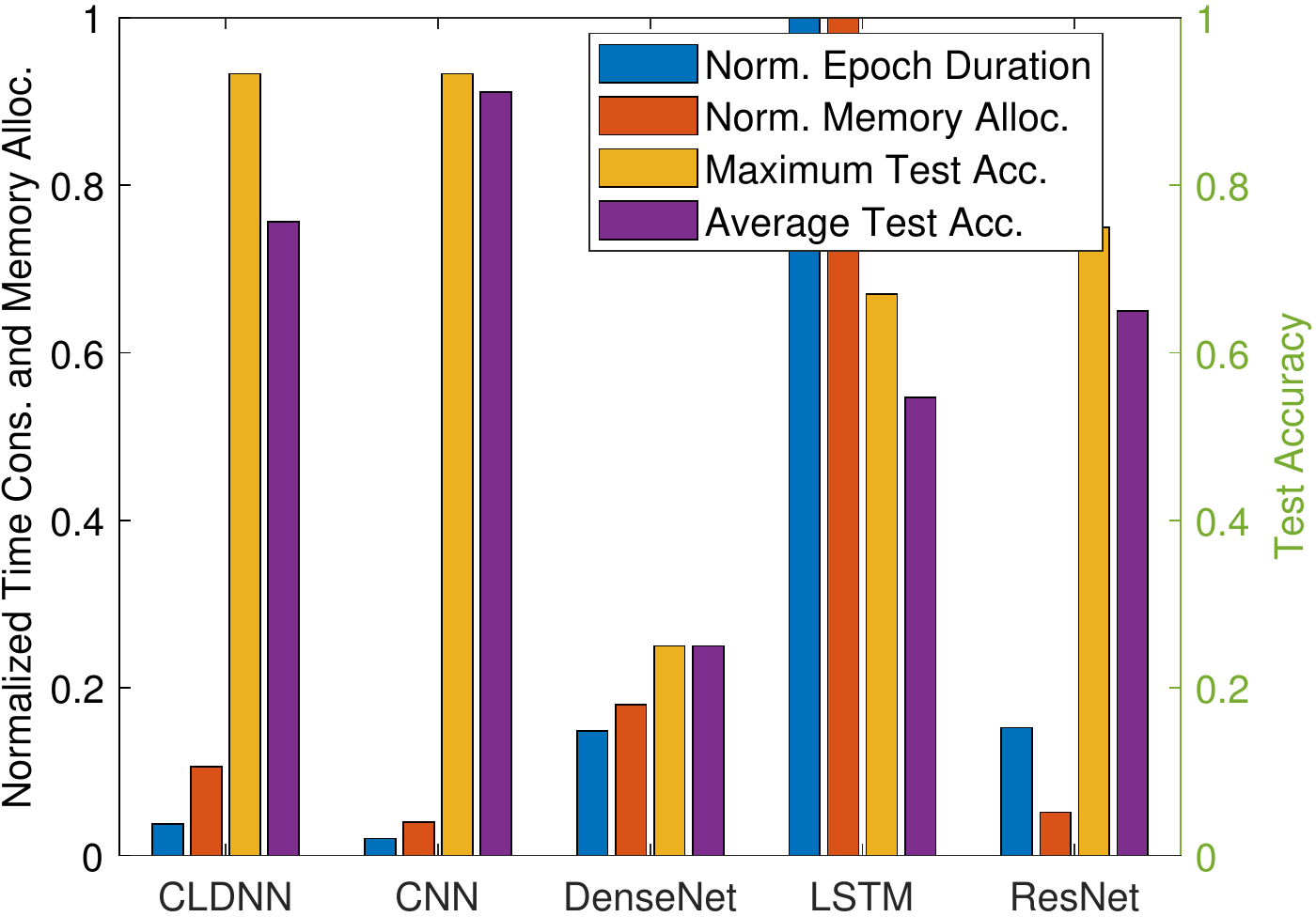}
		\caption{\hl{Model comparison in terms of memory, complexity, and accuracy. The epoch time and memory allocation rate are normalized with their maximum values observed in these models (maximum values for both are observed in \mbox{\ac{LSTM}}). The average accuracy is the mean accuracy in the \mbox{\ac{SNR}} range between 1dB and 15dB.}}
		\label{fig:model_comparison}
	\end{figure}
	
	\begin{figure}[!t]
		\centering
		\includegraphics[width=\linewidth]{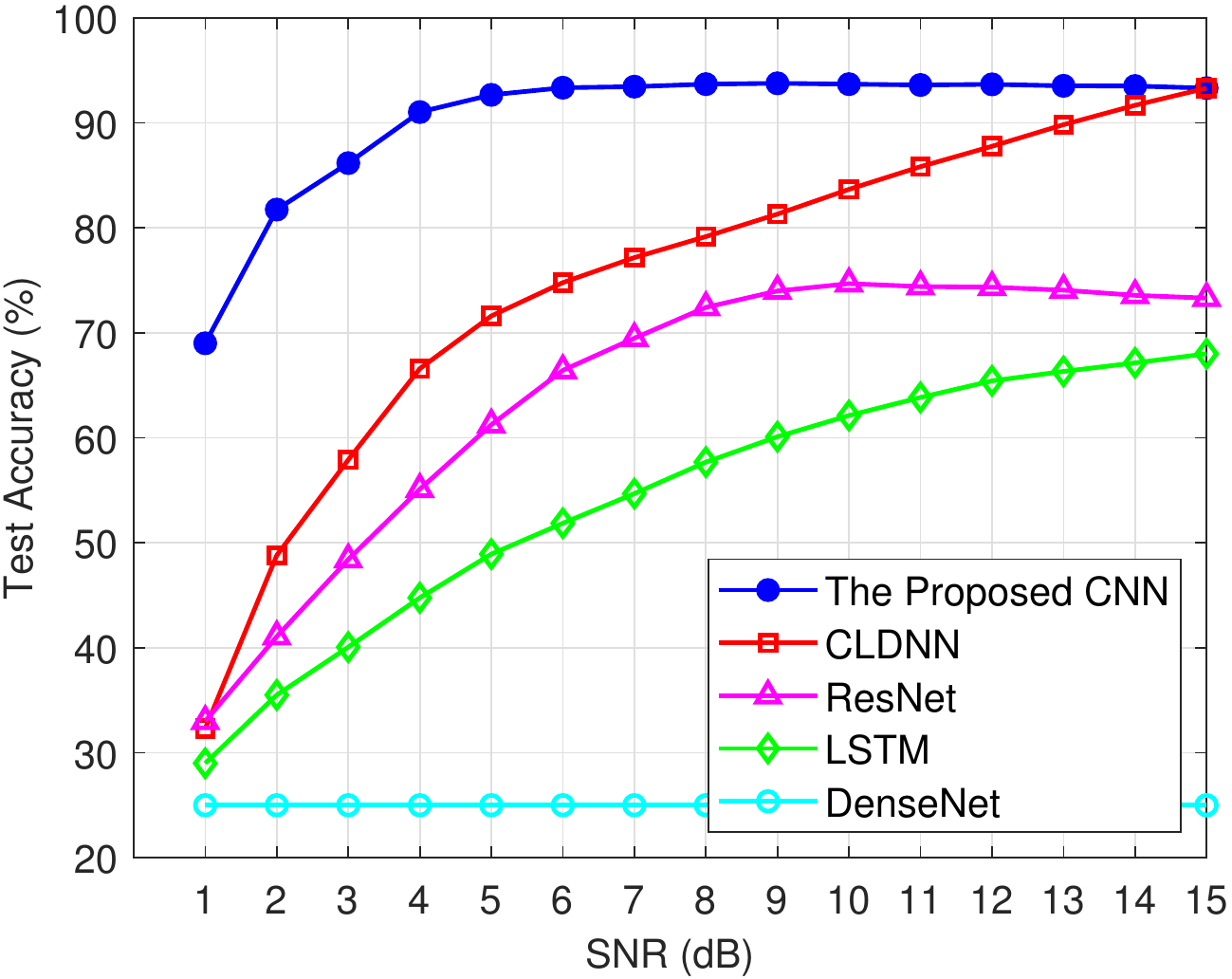}
		\caption{\hl{Test accuracy with respect to \mbox{\ac{SNR}} values for the proposed \mbox{\ac{CNN}}, \mbox{\ac{CLDNN}}, \mbox{\ac{LSTM}}, ResNet, and DenseNet models.}}
		\label{fig:model_acc_comparison}
	\end{figure}

	\hl{As seen in \mbox{\FGR{fig:SCF3}}, the meaningful part of the features is located in the middle of the matrices. In order to investigate the possibility of accuracy improvement and the fair comparison with the existing \mbox{\ac{DL}} networks, we employ a small partition in middle of the \mbox{\ac{SCF}} matrix, where the cyclic characteristics are mainly observed. As stated in \mbox{\SEC{sec:dataset_generation}}, an \mbox{\ac{SCF}} matrix has the dimension $8193\times16$. Therefore, it is not possible to train such a dense model in our server equipped with four NVIDIA Tesla V100 GPUs. To compare our proposed \mbox{\ac{CNN}} architecture with a more dense model, we decrease the dimensions of the \mbox{\ac{SCF}} matrices by using only $16\times16$ part in the middle of the matrices. Only in this way, we are able to train complex models such as \mbox{\ac{LSTM}}~\mbox{\cite{rajendran2018deep}} and DenseNet~\mbox{\cite{huang2017densely}} with \mbox{\ac{SCF}}. Moreover, the proposed \mbox{\ac{CNN}}, \mbox{\ac{CLDNN}}~\mbox{\cite{ramjee2019fast}}, and ResNet~\mbox{\cite{o2018over}} are also trained with the shrunken \mbox{\ac{SCF}} matrices. It is worth saying that, we conduct four-class classification (i.e., \texttt{CASE1}) in this study. The results depicted in \mbox{\FGR{fig:model_comparison}} shows that the proposed \mbox{\ac{CNN}} is favorable in terms of both low complexity (i.e., epoch time) and efficient memory allocation, as well as high test accuracy. During this study, batch sizes are kept same for all models. The memory allocation and training time have been normalized by \mbox{\ac{LSTM}}'s memory allocation rate and training time, respectively; thus, computer-independent results are provided in \mbox{\FGR{fig:model_comparison}}. It should be noted that early stopping is used during training of models and the minimum number of epochs is required by the proposed \mbox{\ac{CNN}}. Furthermore, \mbox{\FGR{fig:model_acc_comparison}} denotes the accuracy with respect to \mbox{\ac{SNR}} levels for each model. By considering results, it can be observed that the proposed \mbox{\ac{CNN}} is more robust and efficient than the existing models. Moreover, it is seen that \mbox{\ac{CNN}} gives better results with this smaller matrix than the complete matrix is used. By eliminating the region except for the meaningful part of \mbox{\ac{SCF}}, the input matrices become more distinct from each other. \mbox{\FGR{fig:SCF3}} implies that \mbox{\ac{SCF}} matrices have similarities except for the meaningful part. The confusion matrices in \mbox{\FGR{fig:sc1_conf_matx_16x16}} for $16\times16$ inputs denote the improvement in the precision of \mbox{\ac{AWGN}}. This explains why the small portion of the matrix can lead to higher accuracy.}
	
	\hl{It is also explored how the dimensions of the small partition affect the performance of the \mbox{\ac{CNN}} model. The results show that using 4 rows does not perform well enough. When using rows between 8 and 128 (as power of two), the results are satisfactory. The test accuracy with respect to input size is demonstrated in \mbox{\FGR{fig:size_comparison}}. It is revealed that by considering the accuracy at lower \mbox{\ac{SNR}} regimes and the training time, $16\times16$ is the most suitable size for the \mbox{\ac{CNN}}.}
	%%%%%%%%%%%%%%%%%%%%%%%%%%%%% Confusion Matrices : Case 1 16x16 %%%%%%%%%%%%%%%%%%%%%%%
	\begin{figure}[!t]
		\centering
		\subfigure[]{
			\label{fig:sc1_conf_matx_1dB_16x16}
			\includegraphics[width=0.47\linewidth]{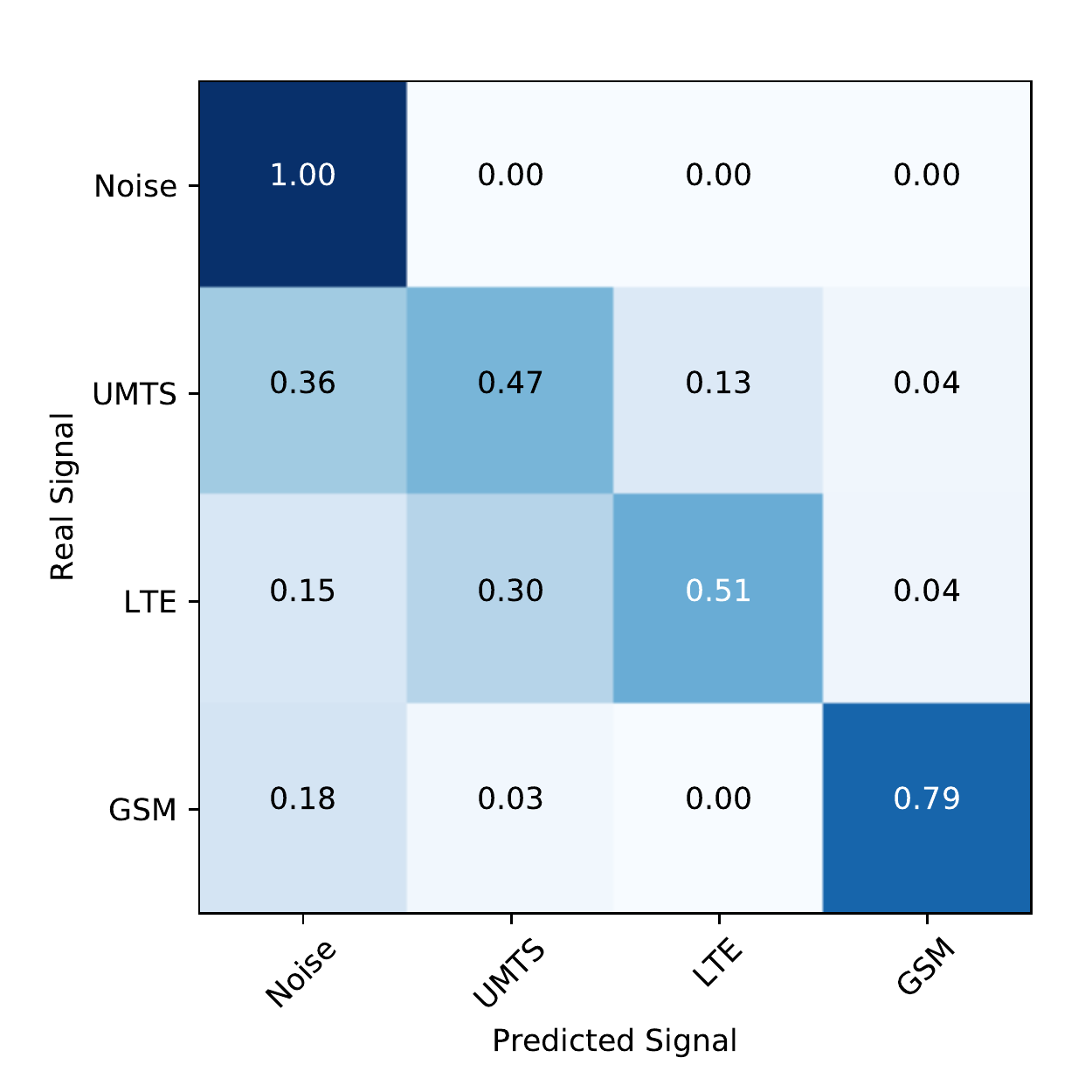}}
		\subfigure[]{
			\label{fig:sc1_conf_matx_5dB_16x16}
			\includegraphics[width=0.47\linewidth]{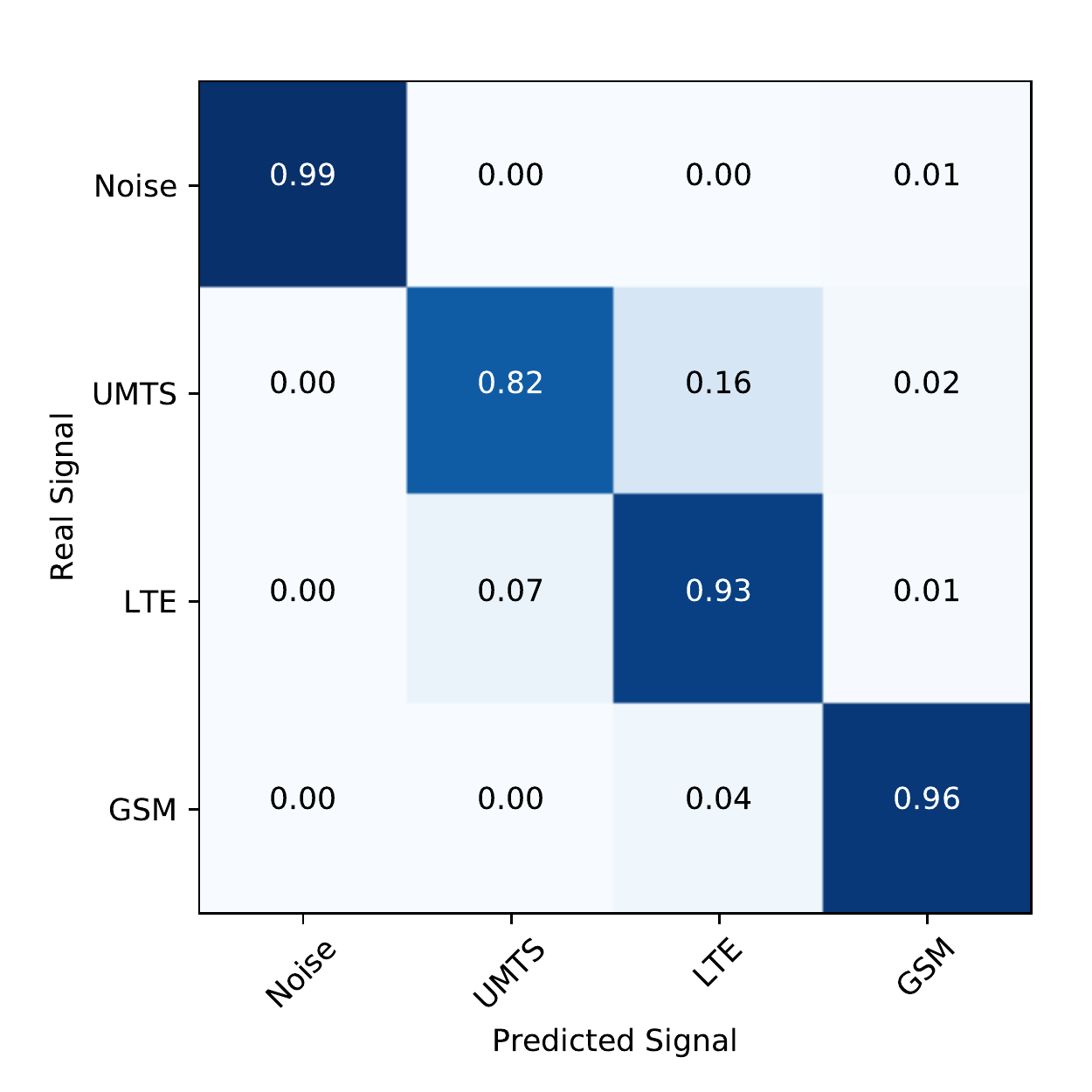}}
		% \subfigure[]{
		% \label{fig:sc1_conf_matx_10dB_16x16}
		% \includegraphics[width=0.32\linewidth]{./figs/results/16x16/10_dB_SCF_Np_4_length_16384_standard_id_snr.pdf}}
		\caption{\hl{Confusion matrices for \texttt{CASE1} at \mbox{\ac{SNR}} levels of (a) 1dB and (b) 5dB when $16\times16$ inputs are employed.}}
		\label{fig:sc1_conf_matx_16x16}
	\end{figure}
	%%%%%%%%%%%%%%%%%%%%%%%%%%%%% Confusion Matrices : Case 1 16x16 %%%%%%%%%%%%%%%%%%%%%%%
	
	\begin{figure}[!t]
		\centering
		\includegraphics[width=\linewidth]{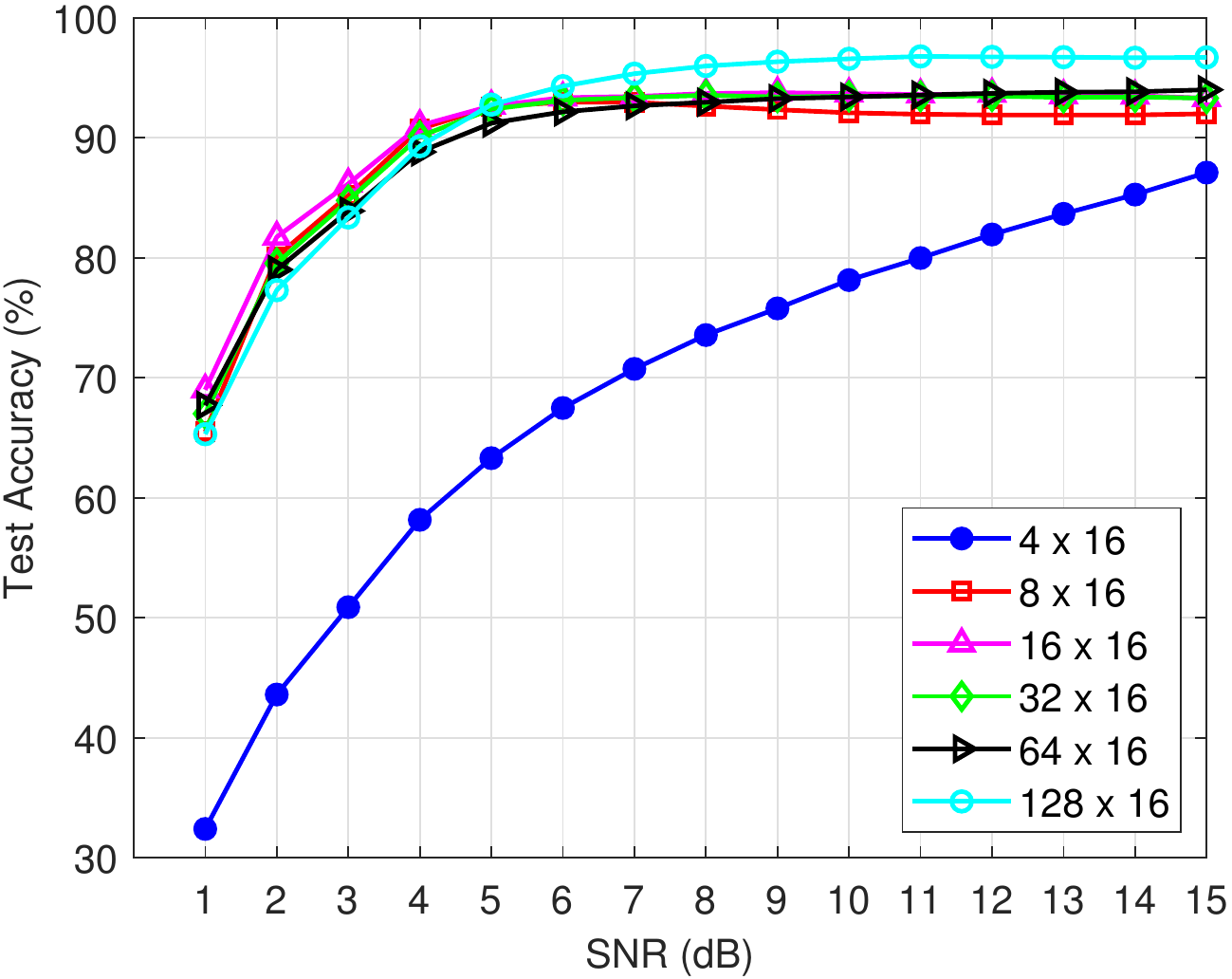}
		\caption{\hl{The test accuracy of the proposed \mbox{\ac{CNN}} architecture with respect to input size.}}
		\label{fig:size_comparison}
	\end{figure}
	
	\section{Conclusion}\label{sec:conclusion}
	In this study, a \ac{DL}-based method utilizing \ac{SCF} as an input to a novel CNN model to achieve spectrum sensing or signal identification interchangeably or jointly without the requirement of any a priori information is proposed. First approach investigates the joint sensing and classification of wireless signals. Second, a sequential approach is adopted. The results show that sequential approach performs better than the joint approach. Moreover, comparative analysis indicated the superiority of \ac{SCF} as a distinctive feature when compare to the contemporary features utilized for currently available \ac{DL}-based detector models. The results also imply that under stringent channel conditions, the \ac{CNN} model of the proposed method provides better spectrum sensing performance than other available \ac{DL} models, SVMs, and classical \ac{CFD}. These results indicate that applicability of \ac{DL}-based techniques in the rapidly changing communications environment of contemporary wireless communications networks. In subsequent studies, the performance of the proposed method for sensing and identification of other wireless signals or modulation techniques with cyclic features can be explored. Furthermore, the performance of the proposed method can be investigated against adversarial attacks and efforts can be made to develop various techniques to strengthen its resistance to these types of intrusions. \textcolor{black}{Although this study focuses on supervised learning, it is possible to improve the performance of the proposed method by supporting unsupervised learning methods in feature extraction.}
	
	\bibliographystyle{IEEEtran}
	\bibliography{tvt_2021}

\begin{thebibliography}{10}
\providecommand{\url}[1]{#1}
\csname url@samestyle\endcsname
\providecommand{\newblock}{\relax}
\providecommand{\bibinfo}[2]{#2}
\providecommand{\BIBentrySTDinterwordspacing}{\spaceskip=0pt\relax}
\providecommand{\BIBentryALTinterwordstretchfactor}{4}
\providecommand{\BIBentryALTinterwordspacing}{\spaceskip=\fontdimen2\font plus
\BIBentryALTinterwordstretchfactor\fontdimen3\font minus
  \fontdimen4\font\relax}
\providecommand{\BIBforeignlanguage}[2]{{%
\expandafter\ifx\csname l@#1\endcsname\relax
\typeout{** WARNING: IEEEtran.bst: No hyphenation pattern has been}%
\typeout{** loaded for the language `#1'. Using the pattern for}%
\typeout{** the default language instead.}%
\else
\language=\csname l@#1\endcsname
\fi
#2}}
\providecommand{\BIBdecl}{\relax}
\BIBdecl

\bibitem{haykin2015cognitive}
S.~Haykin and P.~Setoodeh, ``Cognitive radio networks: The spectrum supply
  chain paradigm,'' \emph{IEEE Trans. Cogn. Commun. Netw.}, vol.~1, no.~1, pp.
  3--28, 2015.

\bibitem{andrews2014will}
J.~G. Andrews, S.~Buzzi, W.~Choi, S.~V. Hanly, A.~Lozano, A.~C. Soong, and
  J.~C. Zhang, ``What will 5g be?'' \emph{IEEE J. Sel. Areas Commun.}, vol.~32,
  no.~6, pp. 1065--1082, 2014.

\bibitem{de2021convergent}
C.~De~Lima, D.~Belot, R.~Berkvens, A.~Bourdoux, D.~Dardari, M.~Guillaud,
  M.~Isomursu, E.-S. Lohan, Y.~Miao, A.~N. Barreto \emph{et~al.}, ``Convergent
  communication, sensing and localization in {6G} systems: An overview of
  technologies, opportunities and challenges,'' \emph{IEEE Access}, 2021.

\bibitem{wild2021joint}
T.~Wild, V.~Braun, and H.~Viswanathan, ``Joint design of communication and
  sensing for beyond {5G} and {6G} systems,'' \emph{IEEE Access}, vol.~9, pp.
  30\,845--30\,857, 2021.

\bibitem{gorcin2014signal}
A.~Gorcin and H.~Arslan, ``Signal identification for adaptive spectrum
  hyperspace access in wireless communications systems,'' \emph{IEEE Commun.
  Mag.}, vol.~52, no.~10, pp. 134--145, 2014.

\bibitem{lees2019deep}
W.~M. Lees, A.~Wunderlich, P.~Jeavons, P.~D. Hale, and M.~R. Souryal, ``Deep
  learning classification of 3.5 {GHz} band spectrograms with applications to
  spectrum sensing,'' \emph{IEEE Trans. Cogn. Commun. Netw.}, 2019.

\bibitem{Qin202020years}
Z.~Qin, X.~Zhou, L.~Zhang, Y.~Gao, Y.-C. Liang, and G.~Y. Li, ``20 years of
  evolution from cognitive to intelligent communications,'' \emph{IEEE Trans.
  Cogn. Commun. Netw.}, vol.~10, no.~1, pp. 6--20, Mar. 2020.

\bibitem{gardner1991}
W.~A. Gardner, ``Exploitation of spectral redundancy in cyclostationary
  signals,'' \emph{IEEE Signal Process. Mag.}, vol.~8, no.~2, pp. 14--36, Apr.
  1991.

\bibitem{o2018over}
T.~J. O'Shea, T.~Roy, and T.~C. Clancy, ``Over-the-air deep learning based
  radio signal classification,'' \emph{IEEE J. Sel. Signal Process.}, vol.~12,
  no.~1, pp. 168--179, Jan. 2018.

\bibitem{kulin2018end}
M.~Kulin, T.~Kazaz, I.~Moerman, and E.~D. Poorter, ``End-to-end learning from
  spectrum data: A deep learning approach for wireless signal identification in
  spectrum monitoring applications,'' \emph{IEEE Access}, vol.~6, pp.
  18\,484--18\,501, Mar. 2018.

\bibitem{kokalj2019autoencoders}
S.~Kokalj-Filipovic, R.~Miller, and J.~Morman, ``{AutoEncoders} for training
  compact deep learning {RF} classifiers for wireless protocols,'' \emph{arXiv
  preprint arXiv:1904.11874}, 2019.

\bibitem{rajendran2018deep}
S.~Rajendran, W.~Meert, D.~Giustiniano, V.~Lenders, and S.~Pollin, ``Deep
  learning models for wireless signal classification with distributed low-cost
  spectrum sensors,'' \emph{IEEE Trans. Cogn. Commun. Netw.}, vol.~4, no.~3,
  pp. 433--445, May. 2018.

\bibitem{han2006spectral}
N.~Han, S.~Shon, J.~H. Chung, and J.~M. Kim, ``{Spectral correlation based
  signal detection method for spectrum sensing in IEEE 802.22 WRAN systems},''
  in \emph{8th International Conference Advanced Communication Technology},
  vol.~3, 2006, pp. 1765--1770.

\bibitem{menguconer_2007}
M.~Oner and F.~Jondral, ``Air interface identification for software radio
  systems,'' \emph{AEU - Intl. J. Electron. Commun.}, vol.~61, no.~2, pp.
  104--117, 2007.

\bibitem{karami2015identification}
E.~Karami, O.~A. Dobre, and N.~Adnani, ``Identification of {GSM} and {LTE}
  signals using their second-order cyclostationarity,'' in \emph{Proc. IEEE
  Int. Instrum. Meas. Tech. Conf.}, Pisa, Italy, May. 2015, pp. 1108--1112.

\bibitem{eldemerdash2017identification}
Y.~A. Eldemerdash, O.~A. Dobre, O.~Ureten, and T.~Yensen, ``Identification of
  cellular networks for intelligent radio measurements,'' \emph{IEEE Trans.
  Instrum. Meas.}, vol.~66, no.~8, pp. 2204--2211, Apr. 2017.

\bibitem{hazza2013overview}
A.~Hazza, M.~Shoaib, S.~A. Alshebeili, and A.~Fahad, ``An overview of
  feature-based methods for digital modulation classification,'' in \emph{Proc.
  Intl. Commun. Signal Process., and Their Applications}, Mar. 2013.

\bibitem{ramjee2019fast}
S.~Ramjee, S.~Ju, D.~Yang, X.~Liu, A.~E. Gamal, and Y.~C. Eldar, ``Fast deep
  learning for automatic modulation classification,'' \emph{arXiv preprint
  arXiv:1901.05850}, 2019.

\bibitem{huang2017densely}
G.~Huang, Z.~Liu, L.~Van Der~Maaten, and K.~Q. Weinberger, ``Densely connected
  convolutional networks,'' in \emph{Proceedings of the IEEE Conference on
  Computer Vision and Pattern Recognition}, 2017, pp. 4700--4708.

\bibitem{safrgh5920}
\BIBentryALTinterwordspacing
K.~Tekb{\i}y{\i}k, {\"O}.~Akbunar, A.~R. Ekti, A.~G{\"o}r{\c{c}}in, and G.~K.
  Kurt, ``{COSINE: Cellular cOmmunication SIgNal datasEt},'' 2020. [Online].
  Available: \url{http://dx.doi.org/10.21227/safr-gh59}
\BIBentrySTDinterwordspacing

\bibitem{roberts1991}
R.~Roberts, W.~Brown, and H.~Loomis, ``Computationally efficient algorithms for
  cyclic spectral analysis,'' \emph{IEEE Signal Process. Mag.}, vol.~8, no.~2,
  pp. 38--49, Apr. 1991.

\bibitem{zhang2017convolutional}
M.~Zhang, M.~Diao, and L.~Guo, ``Convolutional neural networks for automatic
  cognitive radio waveform recognition,'' \emph{IEEE Access}, vol.~5, pp.
  11\,074--11\,082, 2017.

\bibitem{zeiler2013stochastic}
M.~D. Zeiler and R.~Fergus, ``{Stochastic pooling for regularization of deep
  convolutional neural networks},'' \emph{arXiv preprint arXiv:1301.3557},
  2013.

\bibitem{chollet2015keras}
F.~Chollet \emph{et~al.}, ``Keras,'' \url{https://keras.io}, 2015.

\bibitem{cabric2004implementation}
D.~Cabric, S.~M. Mishra, and R.~W. Brodersen, ``Implementation issues in
  spectrum sensing for cognitive radios,'' in \emph{Thirty-Eighth Asilomar
  Conference on Signals, Systems and Computers}, vol.~1, 2004, pp. 772--776.

\bibitem{tekbiyikvtc2020}
K.~{Tekbiyik}, A.~R. {Ekti}, A.~{Gorcin}, G.~K. {Kurt}, and C.~{Kececi},
  ``Robust and fast automatic modulation classification with {CNN} under
  multipath fading channels,'' in \emph{IEEE 91st Vehicular Technology
  Conference (VTC2020-Spring)}, 2020, pp. 1--6.

\bibitem{tekbiyik2019multi}
K.~Tekb{\i}y{\i}k, {\"O}.~Akbunar, A.~R. Ekti, A.~G{\"o}r{\c{c}}in, and G.~K.
  Kurt, ``Multi--dimensional wireless signal identification based on support
  vector machines,'' \emph{IEEE Access}, vol.~7, pp. 138\,890--138\,903, 2019.

\bibitem{hu2018robust}
S.~Hu, Y.~Pei, L.~Pu, and Y.-C. Liang, ``Robust modulation classification under
  uncertain noise condition using recurrent neural network,'' in \emph{{IEEE
  Glob. Commun. Conf.}}, 2018, pp. 1--7.

\bibitem{spooner2018wideband}
C.~M. Spooner and A.~N. Mody, ``Wideband cyclostationary signal processing
  using sparse subsets of narrowband subchannels,'' \emph{IEEE Trans. on Cogn.
  Commun. Netw.}, vol.~4, no.~2, pp. 162--176, 2018.

\end{thebibliography}
	
	%%%%%%%%%%%%%%%}%% BIOGRAPHIES %%%%%%%%%%%%%%%%%%%%%%
	\begin{IEEEbiography}
		[{\includegraphics[width=1in,height=1.25in,clip,keepaspectratio]{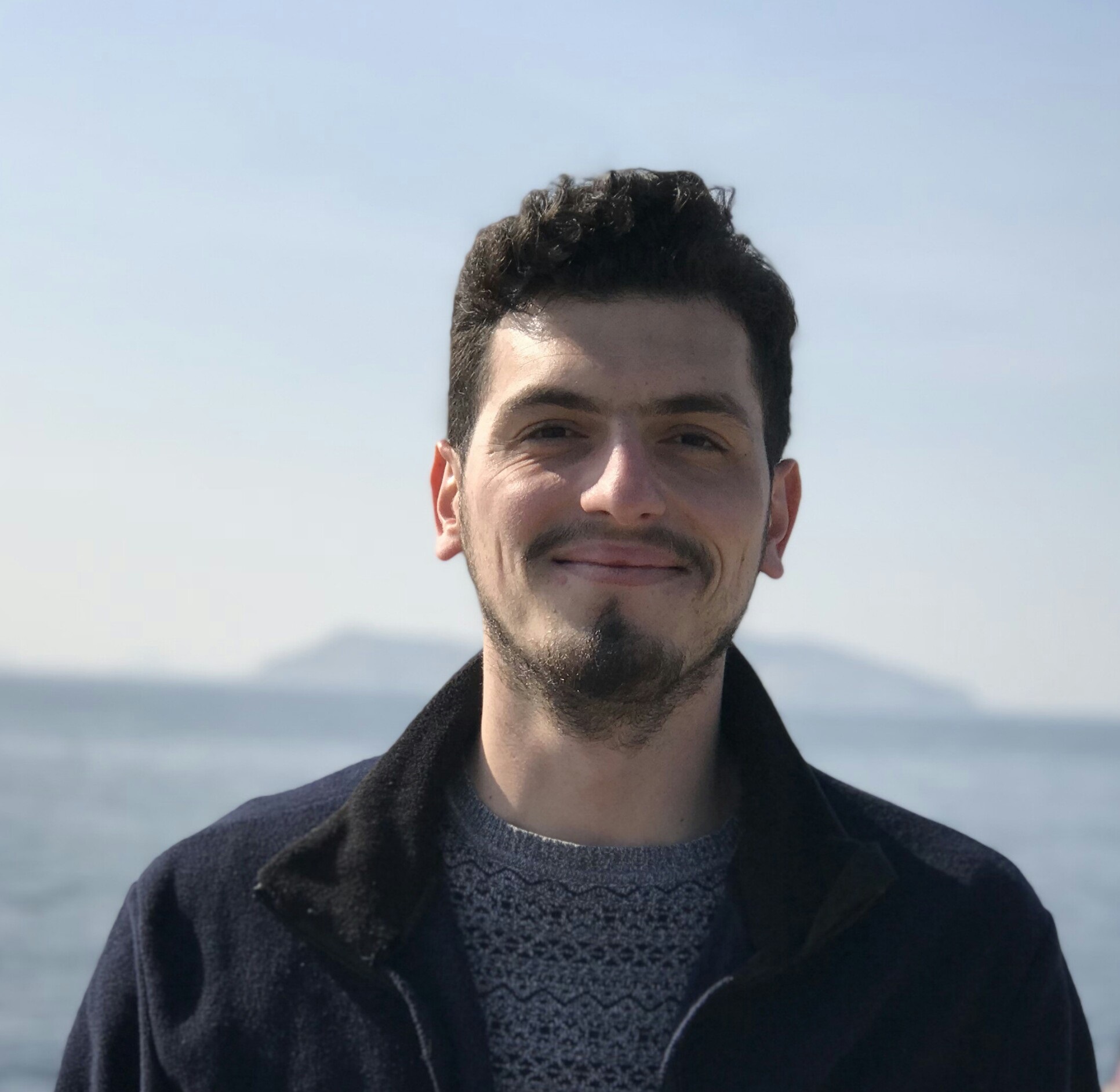}}]{K{\"{u}}r{\c{s}}at Tekb{\i}y{\i}k} received his B.Sc. and M.Sc. degrees (with high honors) in electronics and communication engineering from Istanbul Technical University, Istanbul, Turkey, in 2017 and 2019, respectively. He is currently pursuing his Ph.D. degree in telecommunications engineering in Istanbul Technical University. He is also working as a senior researcher at the HISAR Lab. of TUBITAK BILGEM. His research interests include algorithm design for signal intelligence, next-generation wireless communication systems, terahertz wireless communications, and machine learning.
	\end{IEEEbiography}
	% \vskip
	\begin{IEEEbiography}[{\includegraphics[width=1in,height=1.25in,clip,keepaspectratio]{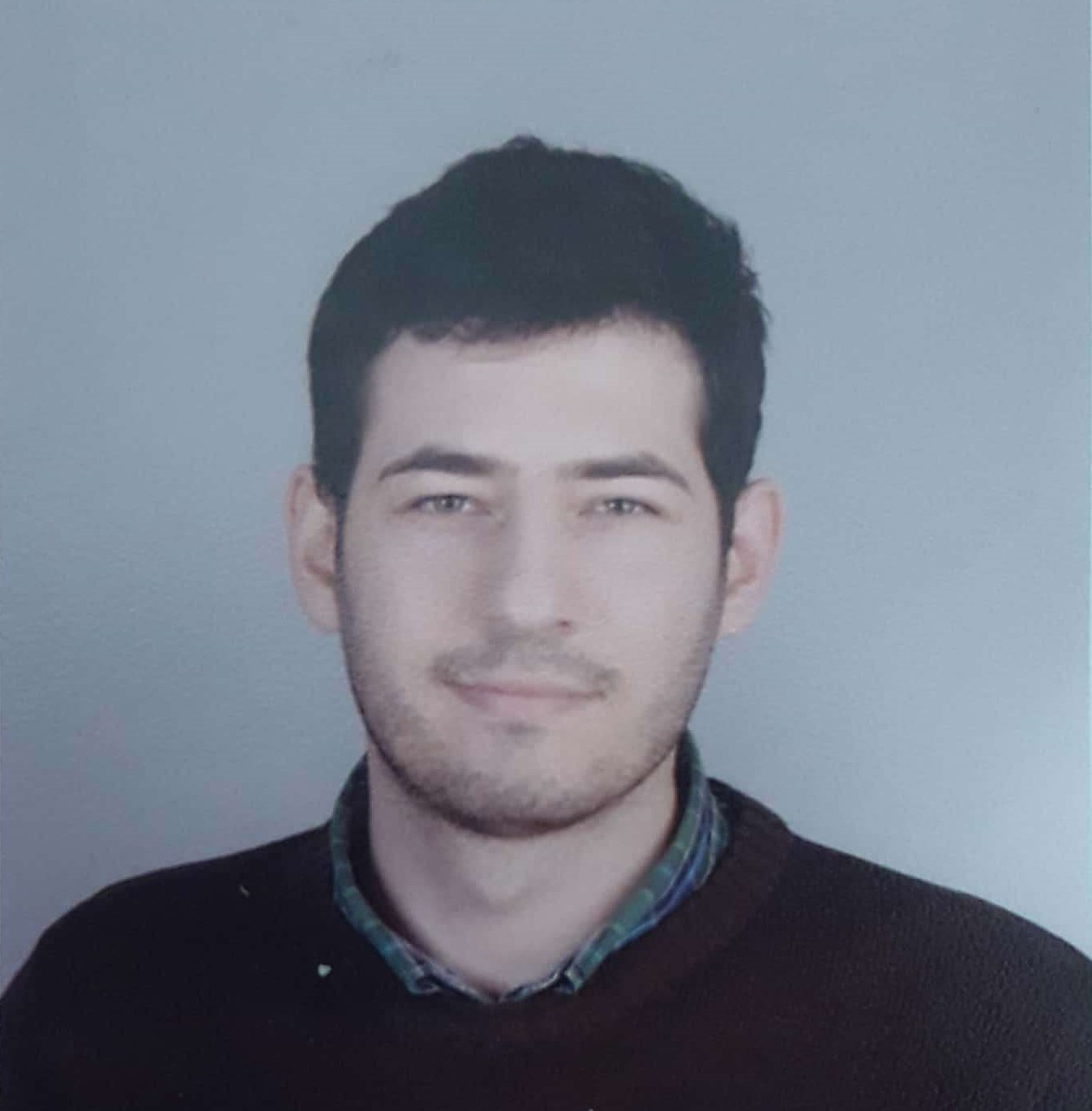}}]{\bf \"{O}zkan Akbunar} received B.Sc. degree from Electrical and Electronics Engineering department in Anadolu University, Eski{\c{s}}ehir, Turkey. He is proceeding his M.Sc education in Telecommunication Engineering department in Istanbul Technical University, Istanbul, Turkey. He is working in the Scientific and Technological Research Council of Turkey at Informatics and Information Security Research Center as R\&D engineer since December 2017. His main research areas are wireless communication systems, digital signal processing, machine learning and radar.
	\end{IEEEbiography}

	\begin{IEEEbiography}
		[{\includegraphics[width=1in,height=1.25in,clip,keepaspectratio]{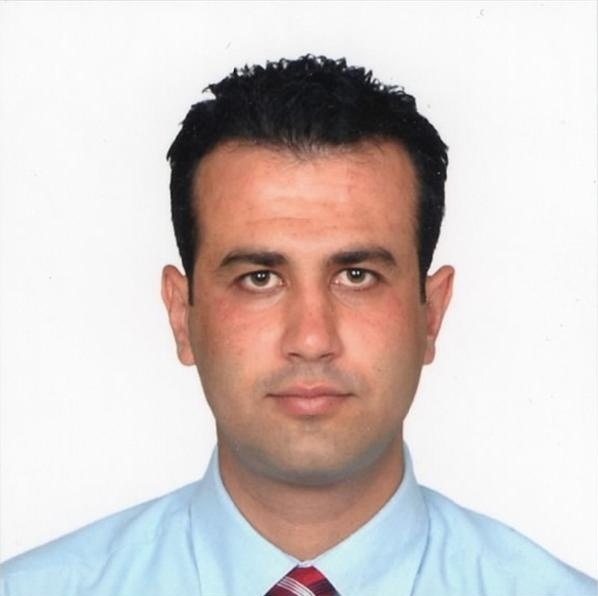}}]{Ali R{\i}za Ekti} is from Tarsus, Turkey. He received B.Sc. degree in Electrical and Electronics Engineering from Mersin University, Mersin, Turkey, (September 2002-June 2006), also studied at Universidad Politechnica de Valencia, Valencia, Spain in 2004-2005, received M.Sc. degree in Electrical Engineering from the University of South Florida, Tampa, Florida (August 2008-December 2009) and received Ph.D. degree in Electrical Engineering from Department of Electrical Engineering and Computer Science at Texas A\&M University (August 2010-August 2015). He is currently an assistant professor at Balikesir University Electrical and Electronics Engineering Department. He also holds the Division Manager position of HISAR Lab. at TUBITAK BILGEM where he is responsible for R\&D activities in the wireless communications and signal processing. His current research interests include statistical signal processing, convex optimization, machine learning, resource allocation and traffic offloading in wireless communications in 5G and beyond systems.
	\end{IEEEbiography}

	\begin{IEEEbiographynophoto}{Ali G\"{o}r\c{c}in} graduated from Istanbul Technical University with B.Sc. in Electronics and
		Telecommunications Engineering and completed his master's degree on defense technologies at the same university. After working at Turkish Science Foundation (TUBITAK) on avionics projects for more
		than six years, he moved to the U.S. to pursue PhD degree in University of South Florida (USF) on wireless communications. He worked for Anritsu Company during his tenure in USF and worked for Reverb Networks and Viavi Solutions after his graduation. He is currently holding an assistant professorship position at Yildiz Technical University in Istanbul and also serving as the interim president of TUBITAK BILGEM.
		%  {\endIEEEbiographynophoto}
	\end{IEEEbiographynophoto}
	
	\begin{IEEEbiography}
		[{\includegraphics[width=1in,height=1.25in,clip,keepaspectratio]{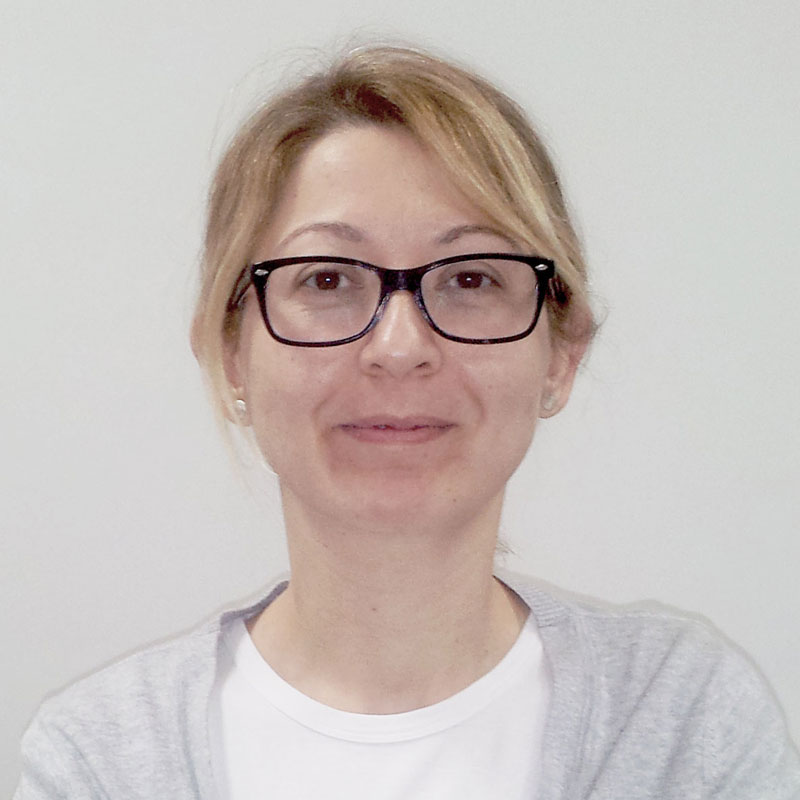}}]{G{\"{u}}ne{\c{s}} Karabulut Kurt} received the B.S. degree with high honors in electronics and electrical engineering from the Bogazici University, Istanbul, Turkey, in 2000 and the M.A.Sc. and the Ph.D. degrees in electrical engineering from the University of Ottawa, ON, Canada, in 2002 and 2006, respectively. From 2000 to 2005, she was a Research Assistant with the CASP Group, University of Ottawa. Between 2005 and 2006, she was with TenXc Wireless, Canada. From 2006 to 2008, she was with Edgewater Computer Systems Inc., Canada. From 2008 to 2010, she was with Turkcell Research and Development Applied Research and Technology, Istanbul. Since 2010, she has been with Istanbul Technical University, where she currently works as a Professor. She is a Marie Curie Fellow. Since August 2019, she is at Carleton University as a visiting professor and also appointed as an adjunct research professor. She is currently serving an Associate Technical Editor  of the \textit{IEEE Communications Magazine}.  
	\end{IEEEbiography}
	
	\begin{IEEEbiography}
		[{\includegraphics[width=1in,height=1.25in,clip,keepaspectratio]{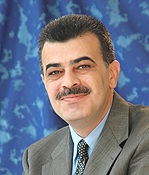}}]{Khalid A. Qaraqe} was born in Bethlehem and received the B.S. (Hons.) degree from the University of Technology, Bagdad, Iraq, in 1986, the M.S. degree from the University of Jordan, Amman, Jordan, in 1989, and the Ph.D. degree in from Texas A\&M University, College Station, TX, in 1997, all in electrical engineering. From 1989 to 2004 he has held a variety positions in many companies and he has over 12 years of experience in the telecommunication industry. He has worked on numerous projects and has experience in product development, design, deployments, testing and integration. He joined the Department of Electrical and Computer Engineering of Texas A\&M University, Qatar, in July 2004, where he is now a Professor and Managing Director with the Center for Remote Healthcare Technology, Qatar. He has been awarded more than 20 research projects consisting of more than USD 13 M from local industries in Qatar and the Qatar National Research Foundation (QNRF). He has published more than 131 journal papers in top IEEE journals, and published and presented 250 papers at prestigious international conferences. He has 20 book chapters published, four books, three patents, and presented several tutorials and talks. His research interests include communication theory and its application to design and performance, analysis of cellular systems and indoor communication systems. Particular interests are in mobile networks, broadband wireless access, cooperative networks, cognitive radio, diversity techniques, index modulation, visible light communication, FSO, telehealth and noninvasive bio sensors.  
	\end{IEEEbiography}
	
	%\balance
\end{document}